\documentclass[revsymb,prb,twocolumn]{revtex4}%
\usepackage{graphicx}
\usepackage{bm}
\usepackage{slashed}
\usepackage{amsmath}
\usepackage{amsfonts}
\usepackage{wasysym}
\usepackage{amssymb}%
\providecommand{\U}[1]{\protect\rule{.1in}{.1in}}
\providecommand{\U}[1]{\protect\rule{.1in}{.1in}}
\newcommand{\be}{\begin{equation}}
\newcommand{\en}{\end{equation}}

\begin{document}
\title{Theory of standing spin waves in finite-size chiral spin soliton lattice}
\author{J. Kishine,$^{1}$ Vl.E. Sinitsyn,$^{2}$ I.G. Bostrem,$^{2}$ Igor
Proskurin,$^{2,3}$ F. J. T. Goncalves,$^{4}$ Y. Togawa$^{4}$ and A.S.
Ovchinnikov,$^{2,5}$}
\affiliation{$^{1}$ Division of Natural and Environmental Sciences, The Open University of
Japan, Chiba 261-8586, Japan}
\affiliation{$^{2}$ Institute of Natural Science, Ural Federal University, Ekaterinburg
620002, Russia}
\affiliation{$3$ Department of Physics and Astronomy, University of Manitoba, Winnipeg,
Manitoba R3T 2N2, Canada}
\affiliation{$4$ Department of Physics and Electronics, Osaka Prefecture University, 1-1
Gakuencho, Sakai, Osaka 599-8531, Japan}
\affiliation{$^{5}$ Institute of Metal Physics, Ural Division, Russian Academy of Sciences,
Ekaterinburg 620219, Russia}
\date{\today }

\begin{abstract}
We present a theory of standing spin wave (SSW) in a monoaxial chiral
helimagnet. Motivated by experimental findings on the magnetic
field-dependence of the resonance frequency in thin films of Cr${}$Nb$_{3}%
$S${}_{6}$[Goncalves et al., Phys. Rev. B \textbf{95}, 104415 (2017)], we
examine the SSW over a chiral soliton lattice (CSL) excited by an ac magnetic
field applied parallel and perpendicular to the chiral axis. For this purpose,
we generalize Kittel-Pincus theories of the SSW in ferromagnetic thin films to
the case of non-collinear helimagnet with the surface end spins which are
softly pinned by an anisotropy field. Consequently, we found there appear two
types of modes. One is a Pincus mode which is composed of a long-period Bloch
wave and a short-period ripple originated from the periodic structure of the
CSL. Another is a short-period Kittel ripple excited by space-periodic
perturbation which exists only in the case where the ac field is applied
perpendicular the chiral axis. We demonstrate that the existence of the Pincus
mode and the Kittel ripple is consistent with experimentally found double
resonance profile.

\end{abstract}

\pacs{Valid PACS appear here}
\maketitle

\section{Introduction}

Dynamical responses to external probes disclose the nature of collective
excitations in condensed matters. Thin ferromagnetic films in this regard have
received much attention in recent years due to striking features never seen in
bulk samples and have been widely applied to technology.\cite{Barman2018}

At the same time, recent studies of thin films and micrometer-sized crystals
of the monoaxial chiral helimagnet CrNb$_{3}$S$_{6}$ have exhibited a number
of remarkable phenomena thereby making this system highly attractive for
spintronic applications. These include detection of the chiral spin soliton
lattice\ (CSL) by using Lorenz microscopy and small-angle electron
diffraction,\cite{Togawa2012} a sequence of jumps in
magnetoresistance\cite{Togawa2013,Wang2017} and magnetic soliton
confinement.\cite{Togawa2015,Book2015} It was argued that the discretization
effects result from a specific domain structure, 1-$\mu$m-wide grains with
different crystallographic structural chirality. In this system, the direction
of in-plane magnetic moments is pinned down around the domain boundary.

An obvious consequence of the pinning effect other than quantization of
magnetization and magneto-resistance is an emergence of the intrinsic
resonance frequency.\cite{Kishine2016} Recent report on magnetic resonance in
micro-sized crystals CrNb$_{3}$S$_{6}$\cite{Goncalves2017} revealed that the
dynamical resonances of the CSL is sensitive to the polarization of the
driving microwave field. In the case where the microwave field is applied
parallel to the chiral axis, the resonance profile was attributed to
excitation of standing spin waves (SSW) \cite{Goncalves2018}. On the other
hand, when the microwave field is applied perpendicular to the chiral axis,
two resonance modes, with the frequency difference being a few GHz, appear
across the entire CSL phase. Furthermore, the resonance modes become
asymmetric with regards to the directions of a static field applied
perpendicular to the chiral axis to stabilize the CSL. Origins of these two
prominent features, (1) double resonance and (2) asymmetry, have not been
known as yet. This situation naturally motivate us to address a query
concerning possible mechanisms.

For this purpose, we start with turning our attention to the theory of
ferromagnetic resonance (FMR) in thin films, pioneered by
Kittel.\cite{Kittel1958} In Kittel's theory, the surface spins are essentially
pinned down by a strong surface anisotropy field. The case of soft pinning was
later considered by Pincus.\cite{Pincus1960} In the case of soft pinning, the
eigenfrequencies of the interior spin wave are required to match the Larmor
frequency of the surface spins. This matching condition is given by
Davis-Puszkarski equation,\cite{Davis1963,Puszkarski1973,Banavar1978} which
leads to allowed values of the wave vector of the spin wave modes.

Since the first observation in permalloy films,\cite{Seavey1958} the detection
of the SSW has long attracted considerable attention, including manganite
films,\cite{Lofland1995} magnonic crystals,\cite{Mruczkiewicz2013}
ferromagnetic bars.\cite{Wismayer2012} The SSW has also been regarded as
candidates for working media in spintronics, including Co
multi-layers,\cite{Buczek2010} a ferrite film,\cite{Serga2007,Chumak2009}
spin-torque excitation in YIG/Co hetero-structures.\cite{Klinger2018}
Detection of the SSW by FMR is also used to probe the interface
exchange-biased structure.\cite{Magaraggia2011}

However, so far little attention has been paid to the SSW in a magnetic system
with a non-collinear ground state, simply because of lack of experimental
motivation. In this regard, we expect that clarifying nature of the SSW\ in
chiral helimagnetic system may open a new window to the field. In particular a
purpose of this paper is to reproduce the magnetic resonance profile in
micro-sized CrNb$_{3}$S$_{6}$,\cite{Goncalves2017} with the aid of a theory of
SSW over the CSL.

This paper is organized as follows. In Sec. II, we describe a model. In Sec.
III, we present results of numerical simulations of SSW based on equations of
motion for the spins. In Sec. IV, we present a detailed analytical theory
based on the generalized Davis-Puszkarski scheme. The discussions and
conclusions are given in Sec. V.

\section{Model}

In this section we present a model to describe the SSW in a mono-axial chiral
helimagnet. What is essential is correct description of the ground state of
finite-size soliton lattice with surface spins at the boundaries of a domain
with a definite chirality. In this respect, we note that in most of the
previous theoretical studies\cite{Book2015} the linear size of the system is
assumed to be infinite, though some confinement effects in a finite size
system has been proposed.\cite{Monchesky2013,Kishine2014}

The pinning effects are implemented through surface anisotropy described by
two equivalent manners. The atomic discrete lattice approach was used in Refs.
\cite{Pincus1960,Soohoo1963} as opposed to continuum model developed by Rado
and Weertman.\cite{Rado1959,Sparks1970} We will follow the former approach below.

\begin{figure}[t]
\begin{center}
\vspace{10mm}
\includegraphics[width=80mm,bb=10 0 974 529]{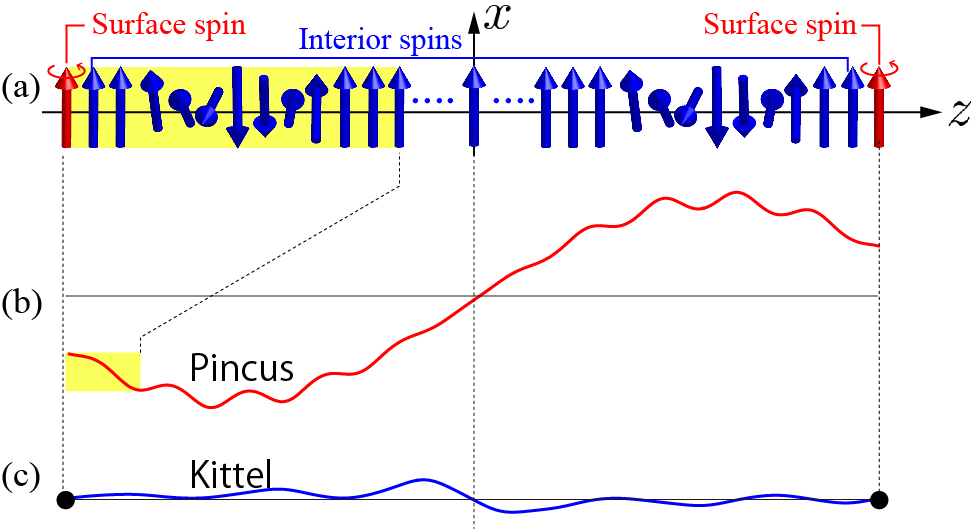}
\end{center}
\caption{(a) The interior (blue) and surface end (red) spins which form the
chiral soliton lattice. (b) The soft boundary condition gives rise to the
Pincus mode which is incommensurate with respect to the background system. (c)
The perpendicular ac field ${\beta}_{x}(\tau)$ further excites the additional
Kittel mode (ripple) which is pinned down at both ends and commensurate with
respect to the background system.}%
\label{schematic_view}%
\end{figure}

CrNb$_{3}$S$_{6}$ crystal has localized spins $S=3/2$ carried by Cr$^{3+}$
ions and the strong intra-layer ferromagnetic coupling strength, $J_{\bot}%
\sim154$K, although the weak inter-layer ferromagnetic coupling strength is
$J\sim18$K and the further weak DM interaction strength is $D\sim
2.9$K.\cite{Shinozaki2016} This layered structure with strong intra-layer
ferromagnetic correlation makes it legitimate to describe the system based on
an effective one-dimensional classical Hamiltonian,%
\begin{align}
H  &  =-J\sum_{\left\langle i,j\right\rangle }\boldsymbol{S}_{i}%
\cdot\boldsymbol{S}_{j}-\boldsymbol{D}\cdot\sum_{\left\langle i,j\right\rangle
}\boldsymbol{S}_{i}\times\boldsymbol{S}_{j}\nonumber\\
&  -\left[  \boldsymbol{H}_{0}+\boldsymbol{h}(t)\right]  \cdot\sum
_{i}\boldsymbol{S}_{i}-\sum_{\sigma=\text{l},\text{r}}\boldsymbol{H}_{s}%
\cdot\boldsymbol{S}_{\sigma}, \label{LatticeHam}%
\end{align}
where $\boldsymbol{S}_{i}$ is the local spin vector located at the site $i$,
$J>0$ is the nearest-neighbor ferromagnetic exchange interaction,
$\boldsymbol{D}=D\hat{\mathbf{e}}_{z}$ is the mono-axial DM interaction vector
along a certain crystallographic chiral axis (taken as the $z$-axis). We take
the $z$-axis as the mono-chiral-axis and let the linear size be $L$. The both
ends are specified by $z=\pm L/2$. $\boldsymbol{H}_{0}=H_{0}\hat{\mathbf{e}%
}_{x}$ is the external magnetic dc-field and $\boldsymbol{h}(t)$ is a
microwave ac-field, given in units of $g\mu_{\text{B}}$. The first two sums
are restricted to the nearest neighbors, the sum over $s$ is a sum over the
spins on the left ($\sigma=$l) and right ($\sigma=$r) boundary surfaces. As
experimentally indicated,\cite{Togawa2015} the constant surface anisotropy
field $\boldsymbol{H}_{s}=H_{\text{s}}\hat{\mathbf{e}}_{x}$ is assumed to lie
in the plane of the film ($xy$-plane), i.e., $\boldsymbol{H}_{s}$ is parallel
to the uniform dc-field $\boldsymbol{H}_{0}$. Because of the soft boundary
condition originated from $\boldsymbol{H}_{s}$, the end surface spins
$\boldsymbol{S}_{\text{l}}$ and $\boldsymbol{S}_{\text{r}}$\ have their own
dynamics distinguished from the interior spins $\boldsymbol{S}_{i}$, as
schematically indicated in Fig. \ref{schematic_view}(a).

Now, we separately treat the dynamics of the interior and end spins. To
analyze the dynamics of the interior spins, the term with $\boldsymbol{H}_{s}$
can be discarded. Long period modulation of the magnetic structure makes it
legitimate to take a continuum limit of the lattice Hamiltonian
(\ref{LatticeHam}),
\begin{align}
\mathcal{H}_{\text{interior}}  &  =\frac{JS^{2}}{2}\left(  \partial_{z}%
\theta\right)  ^{2}+\frac{JS^{2}}{2}\sin^{2}\theta\left(  \partial_{z}%
\varphi\right)  ^{2}\nonumber\\
&  -DS^{2}\sin^{2}\theta\partial_{z}\varphi\nonumber\\
&  -\left[  H_{0}+h_{x}(t)\right]  S\sin\theta\cos\varphi-h_{z}(t)S\cos\theta.
\label{InteriorHam}%
\end{align}
Here, $\varphi$ and $\theta$ are the angles that the magnetization,
$\boldsymbol{S}(\mathbf{r})=S\left(  \sin\theta\cos\varphi,\sin\theta
\sin\varphi,\cos\theta\right)  $, makes with respect to the $x$ and $z$ axis,
respectively, with the film lying parallel to the $xz$ plane. The field
$\boldsymbol{H}_{0}$ is directed along the $x$-axis. The presence of the field
alters a helical spin arrangement of the ground state to the chiral soliton
lattice \cite{Dzyaloshinskii1964,Book2015}. The classical equations of motion,
$\hbar S\sin\theta\partial_{t}\theta=\delta\mathcal{H}_{\text{interior}%
}/\delta\varphi$ and $\hbar S\sin\theta\partial_{t}\varphi=-\delta
\mathcal{H}_{\text{interior}}/\delta\theta$, can be easily shown to be
\begin{align}
\partial_{\tau}\theta &  =-\sin\theta\partial_{z}^{2}\varphi-2\cos
\theta\partial_{z}\theta\partial_{z}\varphi\nonumber\\
&  +2\left(  D/J\right)  \cos\theta\partial_{z}\theta+\left[  \beta_{0}%
+\beta_{x}(t)\right]  \sin\varphi,\label{EOM1}\\
-\sin\theta\partial_{\tau}\varphi &  =\sin\theta\cos\theta\left(  \partial
_{z}\varphi\right)  ^{2}-\partial_{z}^{2}\theta\nonumber\\
&  -2\left(  D/J\right)  \sin\theta\cos\theta\partial_{z}\varphi\nonumber\\
&  -\left[  \beta_{0}+\beta_{x}(t)\right]  \cos\theta\cos\varphi+\beta
_{z}(t)\sin\theta, \label{EOM2}%
\end{align}
where $\tau=JSt/\hbar$, $\beta_{0}=H_{0}/JS$ and $\beta_{x,z}(t)=h_{x,z}%
(t)/JS$. We note that the frequency scale is $JS/h\sim5.6\times10^{11}$Hz (by
choosing $J=18$K\cite{Shinozaki2016} and $S=3/2$).

The ground state is specified by $\theta_{0}=\pi/2$ and%
\begin{equation}
\varphi_{0}(z)=\phi_{0}-2\text{am}\left(  \frac{\sqrt{\beta_{0}}}{\kappa
}z+\phi_{1}\right)  . \label{Ground}%
\end{equation}
Here, $\mathrm{am}\text{ }$is the Jacobi amplitude function. The elliptic
modulus $\kappa$ and two constants,$\ \phi_{0}$($0$ or $\pi$) and $\phi_{1}%
$($0$ or $K$) are chosen through fitting with the numerical data (see Sec.
III). The spatial period of the CSL\ is $L_{\text{CSL}}=2\kappa K/\sqrt
{\beta_{0}}$ and the total number of the solitons is $n=L/L_{\text{CSL}}$.
Here, $K$ is the elliptic integral of the first kind.

To consider small dynamical fluctuations around the equilibrium configuration
of the CSL, we introduce the $\psi\left(  z,t\right)  $ (out-of-plane) and
$\chi\left(  z,t\right)  $ (in-plane) fluctuations of the local spins,%
\begin{equation}
\left\{
\begin{array}
[c]{c}%
\theta(z,\tau)=\dfrac{\pi}{2}+\psi(z,\tau),\\
\varphi(z,\tau)=\varphi_{0}(z)+\chi(z,\tau),
\end{array}
\right.
\end{equation}
where $\left\vert \psi\right\vert ,\left\vert \chi\right\vert \ll1$. Then,
linear approximation of Eqs.(\ref{EOM1}) and (\ref{EOM2}) leads to (see
Appendix A)
\begin{align}
\frac{\kappa^{2}}{\beta_{0}}\frac{\partial\psi\left(  z,t\right)  }%
{\partial\tau}  &  =\hat{\mathcal{L}}_{\varphi}\chi\left(  z,t\right)
+\frac{\kappa^{2}}{\beta_{0}}{\beta}_{x}(\tau)\sin\varphi_{0}\left(  z\right)
,\label{Lame1}\\
\frac{\kappa^{2}}{\beta_{0}}\frac{\partial\chi\left(  z,t\right)  }%
{\partial\tau}  &  =-\hat{\mathcal{L}}_{\theta}\psi\left(  z,t\right)
-\frac{\kappa^{2}}{\beta_{0}}{\beta}_{z}(\tau), \label{Lame2}%
\end{align}
where $\hat{\mathcal{L}}_{\varphi}=-\partial_{\bar{z}}^{2}+2\kappa
^{2}\text{sn}^{2}\bar{z}-\kappa^{2}$ and $\hat{\mathcal{L}}_{\theta}%
=-\partial_{\bar{z}}^{2}+2\kappa^{2}\text{sn}^{2}\bar{z}+4-3\kappa^{2}$ are
the linear Lam\'{e} operators, and $\bar{z}=z\sqrt{\beta_{0}}/\kappa$ with sn
being the Jacobi sn function.

We here comment on the structure of the EOMs. $\hat{\mathcal{L}}_{\varphi}$
and $\hat{\mathcal{L}}_{\theta}$ give the propagating \ wave described by the
eigenfunctions of the Lam\'{e} equation. The propagating solution gives the
spin resonance in an infinite system.\cite{Kishine2009} In the present case,
the soft boundary condition gives rise to standing waves, where the end
surface spins softly fluctuate. We call this mode `Pincus mode' [see Fig.
\ref{schematic_view}(b)] which is incommensurate with respect to the
background system.

Next, the dynamics of the surface spins is described by (see Appendix B for
details)
\begin{align}
\frac{\partial\psi_{s}}{\partial\tau}  &  =\pm\partial_{z}\chi_{s}-\frac{1}%
{2}\partial_{z}^{2}\chi_{s}\mp(D/J)\partial_{z}\varphi_{0}\partial_{z}\chi
_{s}\nonumber\\
&  +{\beta}_{x}(\tau)\sin\varphi_{0}+\left(  \beta_{\text{s}}+\beta
_{0}\right)  \cos\varphi_{0}\cdot\chi_{s},\label{BC1}\\
\frac{\partial\chi_{s}}{\partial\tau}  &  =\mp\partial_{z}\psi_{s}+\frac{1}%
{2}\partial_{z}^{2}\psi_{s}+\frac{1}{2}\left(  \partial_{z}\varphi_{0}\right)
^{2}\psi_{s}\nonumber\\
&  +(D/J)\left(  \partial_{z}\varphi_{0}\right)  \psi_{s}-\frac{1}%
{2}(D/J)\left(  \partial_{z}^{2}\varphi_{0}\right)  \psi_{s}\nonumber\\
&  -\left[  \left(  \beta_{\text{s}}+\beta_{0}\right)  \cos\varphi_{0}\right]
\psi_{s}, \label{BC2}%
\end{align}
where $\beta_{\text{s}}=H_{\text{s}}/JS$ is the pinning field strength. The
upper (lower) sign refers to the end spins at the right (left) end site,
$z=L/2$ or $-L/2$, that indexed by $s=$r or l, respectively.

The precession of the surface spins with the frequency, $\Omega
_{\text{surface}}$, are eventually caught up in the interior spin wave with
the frequency, $\Omega_{\text{interior}}$. Then, the matching condition
$\Omega_{\text{surface}}=\Omega_{\text{interior}}$, which is called the
Davis-Puszkarski equation,\cite{Davis1963,Puszkarski1973,Banavar1978}
determines the overall spin wave dispersion. The SSW modes are quite sensitive
to the direction of the external magnetic $\boldsymbol{h}(t)$, i.e., whether
$\boldsymbol{h}(t)$\ is paralell to the chiral axis [$\boldsymbol{h}%
(t)=h_{0z}\hat{\mathbf{e}}_{z}$] or perpendicular to the chiral axis
[$\boldsymbol{h}(t)=h_{0x}\hat{\mathbf{e}}_{x}$].

Before presenting the detailed analysis, we give an intuitive argument on this
effect. In Eq. (\ref{Lame1}), the term including ${\beta}_{x}(\tau)\sin
\varphi_{0}\left(  z\right)  $ plays a role of space-time dependent external
force whose `spatial frequency' is equal to the spatial period of the CSL,
$L_{\text{CSL}}$. Therefore, when ${\beta}_{x}(\tau)$ is present, there
appears a series of the additional standing spin waves with their basis
spanned by the eigenfunctions of the Lam\'{e} equation. This additional modes
are completely pinned down at the both ends $z=\pm L/2$. We call this mode
`Kittel ripple' [see Fig. \ref{schematic_view}(c)], which appears only for
finite ${\beta}_{x}(\tau)$ and commensurate to the background system. Unlike
the case of ${\beta}_{x}(\tau)$, the term including ${\beta}_{z}(\tau)$ in Eq.
(\ref{Lame2}) is spatially uniform and can excite the Pincus mode only. As we
will discuss in more detail below, the presence or absences of the Kittel
ripple may provide an explanation for the experimentally found difference in
the SSW modes depending on the direction of the ac field.\cite{Goncalves2017}

\section{Numerical simulations}

\subsection{Simulation scheme}

To gain insight into the resonant dynamics, we first perform numerical
simulations similar to those used to study coherent sliding dynamics driven by
crossed magnetic fields in the mono-axial chiral helimagnet \cite{Kishine2012}%
. The numerical analysis is based on the lattice version of Eqs.
(\ref{EOM1},\ref{EOM2}) for the interior spins
\begin{align}
\frac{\partial\theta_{i}}{\partial\tau}  &  =\sqrt{1+\left(  D/J\right)  ^{2}%
}\sin\theta_{i-1}\sin\left(  \varphi_{i}-\varphi_{i-1}-\delta\right)
\nonumber\\
&  -\sqrt{1+\left(  D/J\right)  ^{2}}\sin\theta_{i+1}\sin\left(  \varphi
_{i+1}-\varphi_{i}-\delta\right) \nonumber\\
&  +\left(  \beta_{0}+\beta_{x}\right)  \sin\varphi_{i},
\end{align}%
\begin{align}
\frac{\partial\varphi_{i}}{\partial\tau}  &  =-\left(  \cos\theta_{i+1}%
+\cos\theta_{i-1}\right) \nonumber\\
&  +\sqrt{1+\left(  D/J\right)  ^{2}}\text{cot}\theta_{i}\sin\theta_{i-1}%
\cos\left(  \varphi_{i}-\varphi_{i-1}-\delta\right) \nonumber\\
&  +\sqrt{1+\left(  D/J\right)  ^{2}}\text{cot}\theta_{i}\sin\theta_{i+1}%
\cos\left(  \varphi_{i+1}-\varphi_{i}-\delta\right) \nonumber\\
&  +\left(  \beta_{0}+\beta_{x}\right)  \text{cot}\theta_{i}\cos\varphi
_{i}-\beta_{z},
\end{align}
where $\delta=\text{arctan}(D/J)$, and complemented by equations for the
boundaries (\ref{BCdisc1}) and (\ref{BCdisc2}). For the boundary spins,
$\beta_{x}$ is substituted for $\beta_{\text{s}}$. The length of the system is
chosen $L=411$ which corresponds to the number of kinks accommodated inside
the system, $n_{\text{max}}=[Lq_{s}/2\pi]=10$. The value is about the same
order of magnitude as the number of kinks confined within grains of a definite
crystalline chirality in thin films of CrNb$_{3}$S$_{6}$,\cite{Togawa2012}
where the pitch of the helix is $q_{s}=0.16$.

A search for a static configuration of the ground state is described in detail
in Ref. \cite{Kishine2012}. The static solution found with the aid of the
relaxation method serves as an initial condition for the dynamical equations
addressed by the eight-order Dormand-Prince method with an adaptive step-size
control. To search for a resonant frequency the following procedure is
adopted. Time evolution of $\varphi_{i}(\tau)$, $\theta_{i}(\tau)$ is
determined as a response to the ac-field $\beta_{\alpha}(\tau)=\beta_{\alpha
0}\left[  1-\exp(-\tau)\right]  $ ($\alpha=x,z$) at the start. The Fourier
transform of the ac-field signal is distributed over a continuous frequency
range that enable to localize an approximate position of an intrinsic
resonance. To get its precise value, the spin dynamics equations are again
integrated but now the ac-field is periodical, $\beta_{\alpha}(\tau
)=\beta_{\alpha0}\sin\left(  \Omega\tau\right)  $, where $\alpha=x$ or $z$.
Provided that the frequency $\Omega$ lies near the resonance, this gives rise
to characteristic beatings. Re-examination of these signals by Fourier
analysis specifies an exact position of the resonance frequency.

Here, we separately discuss results of numerical solution for configurations
examined experimentally in Ref. \cite{Goncalves2017}: when (I) the ac
microwave field is applied parallel to the chiral axis, and when (II) the
ac-field is applied perpendicular to the axis.

\subsection{Case I: the ac magnetic field is parallel to the chiral axis}

In the case I, we expect the driving ac-field excites the Pincus modes which
are symmetric with respect to reflection across the center ($z=0$), because
the pinning fields $\beta_{\text{s}}(t)$ act on both ends in a symmetric manner.

\begin{figure}[t]
\vspace{10mm}
\begin{center}
\includegraphics[width=80mm,bb=0 0 777 510]{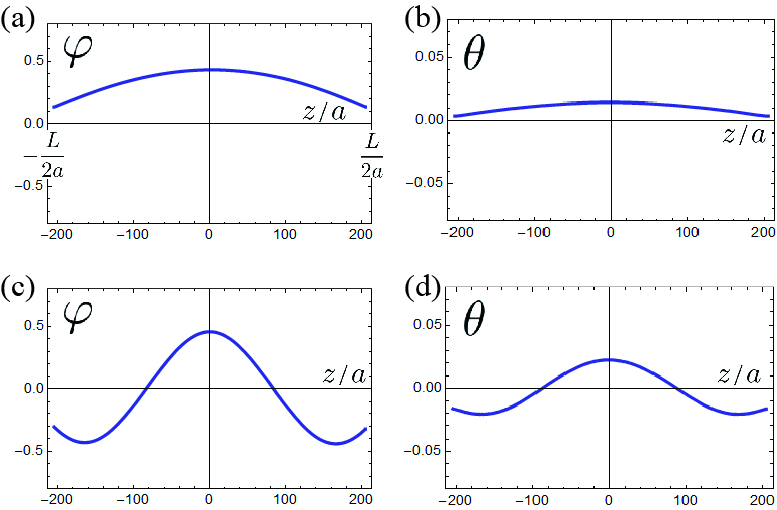}
\end{center}
\caption{The spatial profiles of $\varphi$ and $\theta$ associated with the
standing waves over the helical structure under zero dc magnetic field
($\beta_{0}=0$). The ac field is applied parallel to the chiral axis. The SSW
of the first order at the frequency $\Omega=0.0011$ (a,b) and of the third
order at the frequency $\Omega=0.0030$ (c,d). The parameters are taken as
$\beta_{\text{s}}=0.02$ and $\beta_{z0}=0.0001$ .}%
\label{SW_Parallel}%
\end{figure}

In Fig. \ref{SW_Parallel}, we show the spatial profiles of $\varphi$ and
$\theta$ associated with the standing waves over the helical structure under
zero dc magnetic field ($\beta_{0}=0$). The parameters are taken as
$\beta_{\text{s}}=0.02$ and $\beta_{z0}=0.0001$. The SSW of the first and
third orders occur at the frequencies $\Omega=0.0011$ and $\Omega=0.0030$,
respectively. As evident from Fig. \ref{SW_Parallel}, there arise the SSWs
with the number of half wavelengths being approximately odd in a similar way
to those in the original Kittel's theory.\cite{Kittel1958} It should be noted
that the amplitude of the $\varphi$-oscillations is ten times the $\theta
$-mode amplitude. These findings fit into the idea that spin-wave modes in a
thin film behave like a vibration of a rope clamped at the both ends, when
some anisotropy field essentially or partially pins down boundary spins.

\begin{figure}[t]
\vspace{10mm}
\begin{center}
\includegraphics[width=80mm,bb=0 0 784 533]{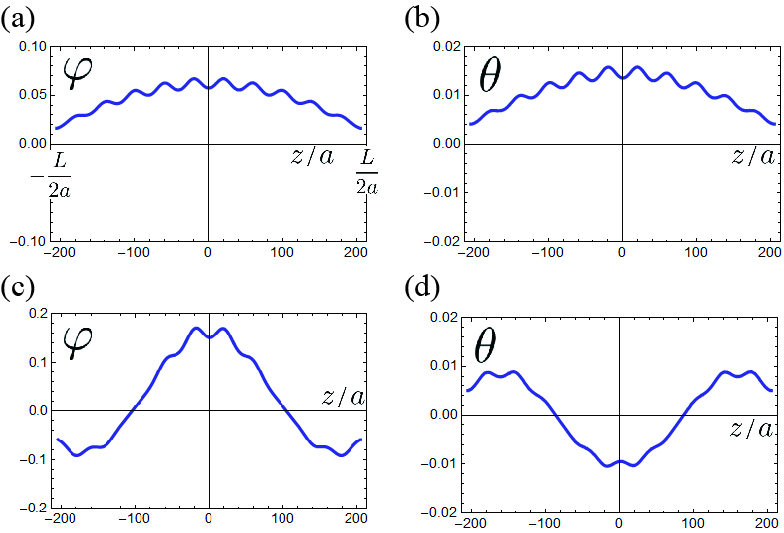}
\end{center}
\caption{The spatial profiles of $\varphi$ and $\theta$ associated with the
standing waves over the CSL structure under finite dc magnetic field
($\beta_{0}=0.002$). The ac field is applied parallel to the chiral axis. The
SSW of the first order at the frequency $\Omega=0.00105$ (a,b), and of the
third order at the frequency $\Omega=0.00305$ (c,d). The parameters are taken
as $\beta_{\text{s}}=0.02$ and $\beta_{x0}=0.0001$.}%
\label{Fig3}%
\end{figure}

In Fig. \ref{Fig3}, we show the spatial profiles of $\varphi$ and $\theta$
associated with the standing waves over the CSL structure under finite dc
magnetic field ($\beta_{0}=0.002$). The parameters are taken as $\beta
_{\text{s}}=0.02$ and $\beta_{x0}=0.0001$. The SSW of the first and third
orders occur at the frequencies $\Omega=0.00105$ and $\Omega=0.00305$,
respectively. Just as in the case of Fig. \ref{SW_Parallel}, one may
immediately recognize the symmetric modes with zero and two nodes, but the
profiles of these excitations look different. It is clear that some
short-scale oscillations, seen as a ripple over the standing wave background,
contribute to the signal. We will discuss the origin of this ripple in Sec.
IV. 

\subsection{Case II: the ac magnetic field is perpendicular to the chiral
axis}

Next we consider the case II. In this case, we expect the driving ac-field
excites the Pincus modes which are antisymmetric with respect to reflection
across the center ($z=0$), because in Eq. (\ref{Lame1}) the space-dependent
field $\sin\varphi_{0}\left(  z\right)  $ is an odd function of $z$.

\begin{figure}[t]
\vspace{10mm}
\begin{center}
\includegraphics[width=80mm,bb=0 0 763 261]{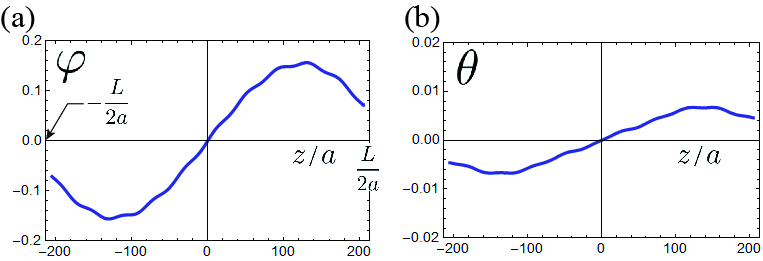}
\end{center}
\caption{The spatial profiles of (a) $\varphi$ and (b) $\theta$ associated
with the second order standing waves over the helical structure under zero dc
magnetic field ($\beta_{0}=0$). The ac field is applied perpendicular to the
chiral axis at the frequency $\Omega=0.0021$.}%
\label{Fig4}%
\end{figure}

In Fig. \ref{Fig4}, we show the spatial profiles of $\varphi$ and $\theta$
associated with the standing waves over the helical structure under a zero dc
magnetic field ($\beta_{0}=0$). The parameters are taken as $\beta_{\text{s}%
}=0.02$ and $\beta_{x0}=0.0001$. The SSW of the second order occurs at the
frequencies $\Omega=0.0021$. As compared with the case I, shown in Fig.
\ref{SW_Parallel}, we recognize that the additional ripples with tiny
amplitude are superimposed on the background Pincus mode. This additional
ripples are caused by the space-time dependent external force, ${\beta}%
_{x}(\tau)\sin\varphi_{0}\left(  z\right)  $. Because the term $\sin
\varphi_{0}\left(  z\right)  $ vanishes at both ends $z=\pm L/2$, the
additional modes are analogous to the original Kittel modes\cite{Kittel1958}
which are completely pinned down at both ends.

\begin{figure}[t]
\vspace{10mm}
\begin{center}
\includegraphics[width=80mm,bb=0 0 763 269]{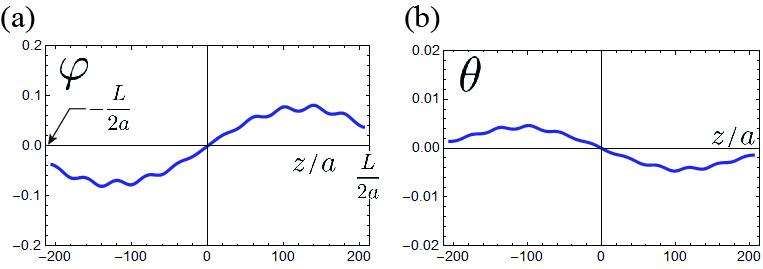}
\end{center}
\caption{The spatial profiles of (a) $\varphi$ and (b) $\theta$ associated
with the second order standing waves over the CSL structure under finite dc
magnetic field ($\beta_{0}=0.002$). The ac field is applied parallel to the
chiral axis at the frequency $\Omega=0.0022$.}%
\label{Fig6new}%
\end{figure}

In Fig. \ref{Fig6new}, we show the spatial profiles of $\varphi$ and $\theta$
associated with the standing waves over the CSL structure under a finite dc
magnetic field ($\beta_{0}=0.002$). The parameters are taken as $\beta
_{\text{s}}=0.02$ and $\beta_{x0}=0.0001$. The SSW of the second order occurs
at the frequencies $\Omega=0.0022$. As with the case shown in Fig. \ref{Fig4},
the additional Kittel ripples are superimposed on the background Pincus mode,
although it is almost invisible because of tiny amplitudes.

Based on the numerical findings presented above, it is evident that the SSW
modes are significantly affected by the movable boundaries which causes Pincus
modes. Furthermore, when the ac field is applied perpendicular to the chiral
axis (case II), the additional Kittel ripples are superimposed. Our next
challenge is to elaborate an appropriate analytical theory of the dynamics.

\section{Analytical theory of the SSW dynamics}

In this section, we present an analytical theory in details. We consider the
following four cases depending on `case I or II' and `the SSW over either
CSL\ or helical structure.' Throughout this section, we follow the technical
scheme for the Davis-Puszkarski equation which is summarized in Appendix C
from general viewpoints.

\subsection{Case I: the ac magnetic field is parallel to the chiral axis}

\subsubsection{SSW over the CSL}

In the case I, the interior spins are subject to the uniform ac-field only. In
Eqs. (\ref{Lame1}) and (\ref{Lame2}), ${\beta}_{x}(\tau)=0$ and only the
spatially uniform field, ${\beta}_{z}(\tau)$, excites the intrinsic SSW modes.
So, we can also drop ${\beta}_{z}(\tau)$ for the purpose of obtaining the SSW
dispersion. Then, the coupled equations are solved by separation of variables,
$\chi(z,\tau)=\nu(\bar{z})Z(\tau)$ and $\psi(z,\tau)=\nu(\bar{z})\xi(\tau)$.
Both $\chi$\ and $\psi$ fields share the same spatial parts $\nu(\bar{z})$
which is the eigenfunction of the Lam\'{e} equation,\cite{Book2015} and
satisfy%
\begin{align}
\hat{\mathcal{L}}_{\varphi}\nu_{q}(\bar{z})  &  =\lambda^{(\varphi)}\nu
_{q}(\bar{z})\nonumber\\
&  ={\kappa}^{\prime2}\text{sn}^{2}(\alpha,\kappa^{\prime})\nu_{q}(\bar
{z}),\label{Lamephi}\\
\hat{\mathcal{L}}_{\theta}\nu_{q}(\bar{z})  &  =\lambda^{(\theta)}\nu_{q}%
(\bar{z})\nonumber\\
&  =\left[  {\kappa}^{\prime2}\text{sn}^{2}(\alpha,\kappa^{\prime}%
)+2+2{\kappa}^{\prime2}\right]  \nu_{q}(\bar{z}). \label{Lametheta}%
\end{align}
Here, the real parameter $\alpha$ lies in the range $-K^{\prime}%
<\alpha<K^{\prime}$, where $K^{\prime}$ is the elliptic integral of the first
kind with the complementary elliptic modulus, ${\kappa}^{\prime2}=1-{\kappa
}^{2}$. The lower index of the eigenfunctions stands for the wave number of
the Bloch wave,%
\begin{equation}
q(\alpha)=\frac{\sqrt{\beta_{0}}}{\kappa}\left[  Z(\alpha,{\kappa}^{\prime
})+\frac{\pi\alpha}{2KK^{^{\prime}}}\right]  , \label{qSL}%
\end{equation}
which is related with the eigenvalues $\lambda^{(\varphi,\theta)}$ through an
implicit parameter $\alpha$. $Z(\alpha,{\kappa}^{\prime})$ represents Jacobi's
zeta-function. The allowed values of $q(\alpha)$ are to be determined by this equation.

The temporal parts $Z(\tau)$ and $\xi(\tau)$ are nothing but the collective
coordinates associate with $\varphi$\ and $\theta$\ fields, respectively, and
describe the collective dynamics of the CSL as a whole. Then, Eqs.
(\ref{Lame1}) and (\ref{Lame2}), become%
\begin{equation}
\left\{
\begin{array}
[c]{c}%
\dot{\xi}(\tau)=C_{1}\left(  \frac{\beta_{0}}{\kappa^{2}}\right)  Z(\tau),\\
\dot{Z}(\tau)=-C_{2}\left(  \frac{\beta_{0}}{\kappa^{2}}\right)  \xi(\tau),
\end{array}
\right.
\end{equation}
and we immediately have the eigenfrequency $\Omega_{0}=(\beta_{0}/\kappa
^{2})\sqrt{C_{1}C_{2}}$ that contains the arbitrary constants $C_{1,2}$ which
characterize the separation of variables [$\left(  \kappa^{2}/\beta
_{0}\right)  [Z(\tau)]^{-1}\partial\xi(\tau)/\partial\tau=[\nu(\bar{z}%
)]^{-1}\hat{\mathcal{L}}_{\varphi}\nu(\bar{z})=C_{1}$, for example].

Using (\ref{Lamephi}) and (\ref{Lametheta}), we have $C_{1}C_{2}%
=\lambda^{(\varphi)}\lambda^{(\theta)}$ which gives rise to the resonance
frequency for the interior SSW,
\begin{equation}
\Omega_{\text{interior}}^{2}\left(  q\right)  =\frac{\beta_{0}^{2}}{\kappa
^{4}}{\kappa}^{\prime2}\text{sn}^{2}(\alpha,\kappa^{^{\prime}})\left[
{\kappa}^{\prime2}\text{sn}^{2}(\alpha,\kappa^{^{\prime}})+2+2{\kappa}%
^{\prime2}\right]  . \label{ResFreqSL}%
\end{equation}
Then, we use the symmetrical solution of the Lam\'{e}
equation,\cite{Book2015}
\begin{equation}
{\nu_{q}}(\bar{z})\propto\Re\left[  \frac{\theta_{4}\left(  \frac{\pi}%
{2K}[\bar{z}-i\alpha-K]\right)  }{\theta_{4}\left(  \frac{\pi\bar{z}}%
{2K}\right)  }e^{-i\bar{q}\bar{z}}\right]  , \label{Bloch}%
\end{equation}
where $\bar{z}=\sqrt{\beta_{0}}/\kappa(z-L/2)+K$ and $\theta_{4}$ is the
Jacobi theta function.

Next, we solve equations of motion for the end surface spins, Eqs.(\ref{BC1})
and (\ref{BC2}), by means of separation of variables, $\chi_{r,l}(z,\tau
)={\nu_{q}}(z)Z_{\text{s}}(\tau)$, $\psi_{r,l}(z,\tau)={\nu_{q}}%
(z)\xi_{\text{s}}(\tau)$. Here, $\bar{q}=(\kappa/\sqrt{\beta_{0}})q(\alpha)$.
It is to be noted that the spatial part ${\nu_{q}}(z)$ is the same as that for
the interior spins. This trick, used throughout this paper, provides the
resonance frequency for the end surface spins,%
\begin{equation}%
\begin{array}
[c]{c}%
\Omega_{\text{surface}}^{2}=\dfrac{1}{\nu_{q}^{2}(\bar{z}_{L/2})}%
\Biggl[\dfrac{\sqrt{\beta_{0}}}{\kappa}\nu_{q}^{^{\prime}}(\bar{z}%
_{L/2})-\dfrac{\beta_{0}}{2\kappa^{2}}\nu_{q}^{^{\prime\prime}}(\bar{z}%
_{L/2})\Biggr.\\
-\dfrac{D}{J}\dfrac{\beta_{0}}{\kappa^{2}}\varphi_{0}^{^{\prime}}(\bar
{z}_{L/2})\nu_{q}^{^{\prime}}(\bar{z}_{L/2})\\
\Biggl.+\left(  \beta_{0}+\beta_{s}\right)  \cos\varphi_{0}(\bar{z}_{L/2}%
)\nu_{q}(\bar{z}_{L/2})\Biggr]\\
\times\Biggl[\dfrac{\sqrt{\beta_{0}}}{\kappa}\nu_{q}^{^{\prime}}(\bar{z}%
_{L/2})-\dfrac{\beta_{0}}{2\kappa^{2}}\nu_{q}^{^{\prime\prime}}(\bar{z}%
_{L/2})\Biggr.\\
-\dfrac{\beta_{0}}{2\kappa^{2}}\nu_{q}(\bar{z}_{L/2})\left(  \varphi
_{0}^{^{\prime}}\right)  ^{2}(\bar{z}_{L/2})\\
-\dfrac{D}{J}\left\{  \dfrac{\sqrt{\beta_{0}}}{\kappa}\varphi_{0}^{^{\prime}%
}(\bar{z}_{L/2})-\dfrac{\beta_{0}}{2\kappa^{2}}\varphi_{0}^{^{\prime\prime}%
}(\bar{z}_{L/2})\right\}  \nu_{q}(\bar{z}_{L/2})\\
\Biggl.+\left(  \beta_{0}+\beta_{\text{s}}\right)  \cos\varphi_{0}(\bar
{z}_{L/2})\nu_{q}(\bar{z}_{L/2})\Biggr].
\end{array}
\label{ResonanceSL}%
\end{equation}
Now, the matching condition (the Davis-Puszkarski equation)%
\begin{equation}
\Omega_{\text{interior}}\left(  q\right)  =\Omega_{\text{surface}}
\label{matching-condition}%
\end{equation}
leads to the determination of the parameter $\alpha$, and then gives the
allowed wavenumber $q$. This algorithm is similar to the case for
ferromagnetic thin films \cite{Davis1963,Puszkarski1973,Banavar1978}.

\begin{figure}[t]
\vspace{10mm}
\begin{center}
\includegraphics[width=60mm,bb=0 0 382 603]{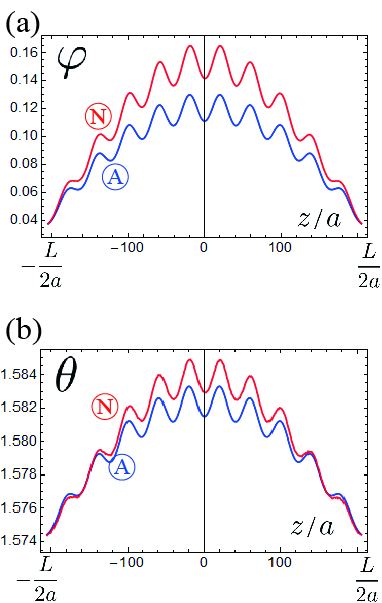}
\end{center}
\caption{Comparison between numerical (red line) and analytical (blue line)
results for the spatial profiles of (a) $\varphi$ and (b) $\theta$ associated
with the first order standing waves over the CSL structure under finite dc
magnetic field ($\beta_{0}=0.002$). The parameters are taken as $\beta
_{\text{s}}=0.02$ and $\beta_{x0}=0.0001$. Analytical result is obtained by
Eqs. (\ref{SymRipple}) at $\tau=5600$. }%
\label{Fig6}%
\end{figure}

The SSW is obtained by superposition of two waves propagating into the
opposite directions,%
\begin{equation}
\chi(\bar{z},\tau)=\frac{{\nu_{q}}(\bar{z}_{L/2+z})+{\nu_{q}}(\bar{z}%
_{L/2-z})}{{\nu_{q}}(\bar{z}_{L})+{\nu_{q}}(\bar{z}_{0})}\chi(\bar{z}%
_{L/2},\tau). \label{SymRipple}%
\end{equation}
We obtain a similar expression for $\psi(\bar{z},\tau)$. Here, the boundary
functions are taken from the numerical data. In Fig. \ref{Fig6}, we show
comparison between numerical and analytical results for the spatial profiles
of $\varphi$ and (b) $\theta$ associated with the first order standing waves
over the CSL structure under finite dc magnetic field ($\beta_{0}=0.002$). It
is seen that the analytical results are consistent with the numerical ones.

\begin{figure}[t]
\vspace{10mm}
\begin{center}
\includegraphics[width=65mm,bb=0 0 451 430]{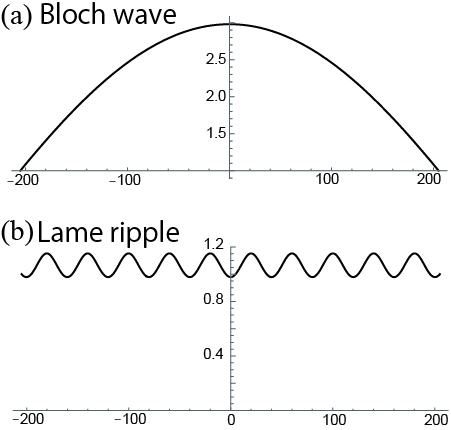}
\end{center}
\caption{Schematic demonstration of how the SSW wave function over the CSL,
Eq. (\ref{Bloch}), is decomposed into (a) Bloch wave ($e^{-i\bar{q}\bar{z}}$)
and (b) Lam\'{e} ripple [ ${\theta_{4}\left(  \frac{\pi}{2K}[\bar{z}%
-i\alpha-K]\right)  }/{\theta_{4}\left(  \frac{\pi\bar{z}}{2K}\right)  }$].
The parameter setting is the same as in the case of Fig. \ref{Fig6}.}%
\label{Fig6b}%
\end{figure}

Analytical results enable us to understand the origin of the
ripples. For this purpose, we decompose Eq. (\ref{Bloch}) into `Bloch wave'
($e^{-i\bar{q}\bar{z}}$) and `Lam\'{e} ripple' [ ${\theta_{4}\left(  \frac
{\pi}{2K}[\bar{z}-i\alpha-K]\right)  }/{\theta_{4}\left(  \frac{\pi\bar{z}%
}{2K}\right)  }$]. We separately show the spatial profiles of these waves in
Fig. \ref{Fig6b}. It is seen that the Bloch part behaves like a smooth
background, while the Lam\'{e} part exhibits the short-wavelength modulation
(ripple) which directly reflects the spatial period of the CSL, $L_{\text{CSL}%
}$.

\begin{figure}[t]
\vspace{10mm}
\begin{center}
\includegraphics[width=70mm,bb=0 0 530 469]{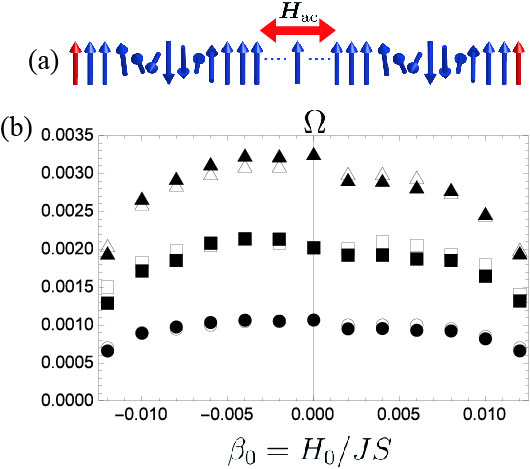}
\end{center}
\caption{(a) Schematic view of the case I configuration. (b) Resonance
frequency $\Omega_{0}$ for standing waves of the first order ($\Circle$ -
numerical data, $\CIRCLE$ - theory), the second order ($\Box$ - numerical
data, $\blacksquare$ - theory) and the third order ($\vartriangle$ - numerical
data, $\blacktriangle$ - theory) depending on the transverse dc-field
$\beta_{0}$.}%
\label{Fig7}%
\end{figure}

In Fig. \ref{Fig7}, the spectrum of the SSW of the first, second and third
orders, where both numerical and analytical results are shown for comparison.
It is evident that the theoretical and numerical results are in good agreement.

\subsubsection{SSW over the simple helix}

We here separately discuss the case of zero magnetic dc-field ($\beta_{0}=0$).
In this case, the ground state is a simple helix with an uniform modulation
and the above analysis can be repeated in much simpler manner.

The helical structure of the interior spins is described by $\theta_{0}=\pi/2$
and $\varphi_{0}=q_{s}(z-z_{0})$. The helical pitch is determined through an
equation (see Appendix B for derivation)
\begin{equation}
0=\sin q_{s}-\left(  D/J\right)  \cos q_{s}\mp\beta_{\text{s}}\sin
\varphi\left(  \mp L/2\right)  . \label{stat_right}%
\end{equation}
The presence of the pinning field complicate the condition for $q_{s}$.

The equations of motion (\ref{Lame1}) and (\ref{Lame2}) may be recast in the
form
\begin{align}
\partial_{\tau}\psi &  =-\partial_{z}^{2}\chi,\\
\partial_{\tau}\chi &  =\left[  \partial_{z}^{2}-2({D}/{J})q_{s}+q_{s}%
^{2}\right]  \psi-\beta_{z}(\tau).
\end{align}
We look for solutions in the form of the standing wave, $\cos(kz)$, with the
wavevector $k$. The resonance frequency for the interior spins is then given
by
\begin{equation}
\Omega_{\text{interior}}^{2}=k^{2}\left[  k^{2}-q_{s}^{2}+2q_{s}({D}%
/{J})\right]  . \label{FreqSpiral}%
\end{equation}
On the other hand, the resonance frequency for the surface end spins is
obtained to be%
\begin{align}
\Omega_{\text{surface}}^{2}  &  =\left[  k\tan\left(  k{L/2}\right)  \left\{
1+(D/J)q_{s}\right\}  \right. \nonumber\\
&  \left.  -k^{2}/2-\beta_{\text{s}}\cos\varphi_{0}\left(  {L/2}\right)
\right] \nonumber\\
&  \times\left\{  k\tan\left(  k{L/2}\right)  -k^{2}/2+q_{s}^{2}/2\right.
\nonumber\\
&  \left.  -(D/J)q_{s}-\beta_{\text{s}}\cos\varphi_{0}\left(  {L/2}\right)
\right\}  . \label{FinalEven1}%
\end{align}
Then, the Davis-Puszkarski equation leads to the determination of allowed
values of $k$.

By performing numerical estimations for $D/J=0.16$, $\beta_{\text{s}}=0.02$,
$\varphi_{0}(\mp L/2)=\pm31.6369$, where $L/2=205$, we obtain $k^{(1)}%
=0.006671$, $\Omega_{0}^{(1)}=0.001067$ and $k^{(3)}=0.020279$, $\Omega
_{0}^{(3)}=0.003268$ for the standing waves of the 1st and 3rd orders,
respectively\cite{Termin}. This result is included in Fig. \ref{Fig7}.

\subsection{Case II: the ac magnetic field is perpendicular to the chiral
axis}

\subsubsection{SSW over the CSL}

Now we examine the most complicated case, where the SSW is excited over the
CSL by the ac-field applied perpendicular to the chiral axis.\ An essential
feature of this case lies in the fact that the interior spins experience
space-time dependent Zeeman interaction with an effective field $\beta
_{\text{eff}}(z,\tau)=\sin\varphi_{0}(z){\beta}_{x}(\tau)$ in Eqs.
(\ref{Lame1}). This term prevents us to apply separation of variables. To
attack the problem the method of transformations leading to homogeneous
boundary conditions may be used \cite{Polyanin2002}. For the purpose, we solve
initially the EOM for the interior spins, by assuming that dynamics of the end
spins is known. The results obtained in this fashion are then used to address
the problem to describe dynamics of the end surface spins in a self-consistent manner.

To realize the scheme the spin fluctuations may be expanded as
\begin{align}
\chi(\bar{z},\tau)  &  =\frac{\nu_{q}(\bar{z})}{\nu_{q}(\bar{z}_{L/2})}%
\chi(\bar{z}_{L/2},\tau)+\epsilon\tilde{\chi}(\bar{z},\tau),\label{Split1}\\
\psi(\bar{z},\tau)  &  =\frac{\nu_{q}(\bar{z})}{\nu_{q}(\bar{z}_{L/2})}%
\psi(\bar{z}_{L/2},\tau)+\epsilon\tilde{\psi}(\bar{z},\tau), \label{Split2}%
\end{align}
where the first terms correspond to the standing wave antisymmetric with
respect to reflection across the center,
\begin{equation}
{\nu_{q}}(\bar{z})\propto\Im\left[  \frac{\theta_{4}\left(  \frac{\pi}%
{2K}[\bar{z}-i\alpha-K]\right)  }{\theta_{4}\left(  \frac{\pi\bar{z}}%
{2K}\right)  }e^{-i\bar{q}\bar{z}}\right]  . \label{BlochA}%
\end{equation}
The second terms in the r.h.s. of Eqs. (\ref{Split1}) and (\ref{Split2}) give
rise to the additional short-wavelength ripples with the both ends being
completely pinned, i.e., $\tilde{\chi}(\bar{z}_{\pm L/2},\tau)=0$ and
$\tilde{\psi}(\bar{z}_{\pm L/2},\tau)=0$. In view of the complete pinning, we
call this additional ripple the `Kittel ripple.'

The amplitude of the Kittel ripple is proportional to the small parameter
$\epsilon$. From the beginning, the dynamics of the end spins, i.e. $\chi
(\bar{z}_{L/2},\tau)$ and $\psi(\bar{z}_{L/2},\tau)$, is considered as being
known. Due to odd parity of the function $\nu_{q}(\bar{z})$, the oscillations
of the boundary spins are anti-synchronized, i.e. $\chi(\bar{z}_{-L/2}%
,\tau)=-\chi(\bar{z}_{L/2},\tau)$ and $\psi(\bar{z}_{-L/2},\tau)=-\psi(\bar
{z}_{L/2},\tau)$.

Substituting Eqs. (\ref{Split1}) and (\ref{Split2}) into Eqs.(\ref{Lame1}) and
(\ref{Lame2}), we obtain the coupled equations of the zeroth-order in
$\epsilon$,%
\begin{align}
\dot{\psi}\left(  \bar{z}_{L/2},\tau\right)   &  =\left(  \beta_{0}/\kappa
^{2}\right)  \lambda^{(\varphi)}\chi\left(  \bar{z}_{L/2},\tau\right)  ,\\
\dot{\chi}\left(  \bar{z}_{L/2},\tau\right)   &  =-\left(  \beta_{0}%
/\kappa^{2}\right)  \lambda^{(\theta)}\psi\left(  \bar{z}_{L/2},\tau\right)  ,
\end{align}
which gives the resonance frequency for the Pincus mode of the interior
spins,
\begin{gather}
\Omega_{\text{interior-Pincus}}^{2}=\frac{\beta_{0}^{2}}{\kappa^{4}}%
\lambda^{(\varphi)}\lambda^{(\theta)}\nonumber\\
=\frac{\beta_{0}^{2}}{\kappa^{4}}{\kappa}^{\prime2}\text{sn}^{2}(\alpha
,\kappa^{^{\prime}})\left[  {\kappa}^{\prime2}\text{sn}^{2}(\alpha
,\kappa^{^{\prime}})+2+2{\kappa}^{\prime2}\right]  , \label{Omega0}%
\end{gather}
with no restriction on $\alpha$. We call this `interior Pincus' mode.

The coupled equations of the first-order in $\epsilon$ are found to be
\begin{align}
\dot{\tilde{\psi}}(\bar{z},t)  &  =\frac{\beta_{0}}{\kappa^{2}}\hat
{\mathcal{L}}_{\varphi}\tilde{\chi}(\bar{z},t)+\frac{\beta_{x}(\tau)}%
{\epsilon}\sin\varphi_{0}(\bar{z}),\label{FirstOrder}\\
\dot{\tilde{\chi}}(\bar{z},t)  &  =-\frac{\beta_{0}}{\kappa^{2}}%
\hat{\mathcal{L}}_{\theta}\tilde{\psi}(\bar{z},t). \label{SecondOrder}%
\end{align}
The resonance frequency for the Kittel ripples of the interior spins is now%
\begin{align}
\Omega_{\text{interior-Kittel}}^{2}\left(  q\right)   &  =\frac{\beta_{0}^{2}%
}{\kappa^{4}}{\kappa}^{\prime2}\text{sn}^{2}(\alpha,\kappa^{^{\prime}%
})\nonumber\\
&  \times\left[  {\kappa}^{\prime2}\text{sn}^{2}(\alpha,\kappa^{^{\prime}%
})+2+2{\kappa}^{\prime2}\right]  , \label{Interior-Kittel-Omega}%
\end{align}
under the restriction on $\alpha$ due to the boundary condition, $\nu_{q}%
(\bar{z}_{\pm L/2})=0$. We look for solutions of Eqs.(\ref{FirstOrder}) and
(\ref{SecondOrder}) in the form
\begin{align}
\tilde{\psi}(\bar{z},\tau)  &  =\sum_{n}\tilde{\psi}_{n}(\tau)\nu_{q_{n}}%
(\bar{z}),\label{FTfluct2}\\
\tilde{\chi}(\bar{z},\tau)  &  =\sum_{n}\tilde{\chi}_{n}(\tau)\nu_{q_{n}}%
(\bar{z}), \label{FTfluct1}%
\end{align}
where the wavevectors $q_{n}$ are determined by $\nu_{q_{n}}(\bar{z}_{\pm
L/2})=0$.

Inserting Eqs. (\ref{FTfluct2}) and (\ref{FTfluct1}) into Eqs.
(\ref{FirstOrder}) and (\ref{SecondOrder}) yields
\begin{align}
\tilde{\chi}_{n}(\tau)  &  =f_{n}\frac{\beta_{x0}}{\epsilon}\frac{\kappa^{2}%
}{\beta_{0}}\frac{\Omega_{n}}{\lambda_{n}^{(\varphi)}}\frac{\left[  \Omega
_{n}\sin(\Omega\tau)-\Omega\sin(\Omega_{n}\tau)\right]  }{\Omega^{2}%
-\Omega_{n}^{2}},\label{BarChi}\\
\tilde{\psi}_{n}(\tau)  &  =f_{n}\frac{\beta_{x0}}{\epsilon}\Omega
\frac{\left[  \cos(\Omega_{n}\tau)-\cos(\Omega\tau)\right]  }{\Omega
^{2}-\Omega_{n}^{2}}, \label{BarPsi}%
\end{align}
where $\Omega_{n}$ is given by (\ref{Interior-Kittel-Omega}) being estimated
for the $q_{n}$, and
\begin{equation}
f_{n}=\frac{\int_{-\bar{z}_{L/2}}^{\bar{z}_{L/2}}\sin\varphi_{0}(\bar{z}%
)\nu_{q_{n}}(\bar{z})d\bar{z}}{\int_{-\bar{z}_{L/2}}^{\bar{z}_{L/2}}\nu
_{q_{n}}^{2}(\bar{z})d\bar{z}}.
\end{equation}
The function $\sin\varphi_{0}$ is antisymmetric with respect to reflection
across the center of the system, and therefore the summation in
(\ref{FTfluct2}) and (\ref{FTfluct1}) should include the odd functions
$\nu_{q_{n}}(\bar{z})$ only.

\begin{figure}[t]
\vspace{10mm}
\begin{center}
\includegraphics[width=70mm,bb=0 0 370 599]{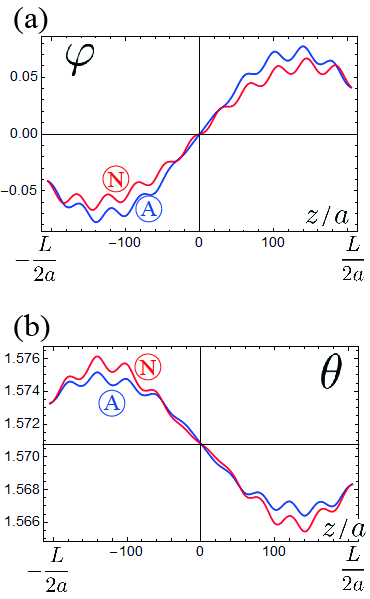}
\end{center}
\caption{ Comparison between numerical (red line) and analytical (blue line)
results for the spatial profiles of (a) $\varphi$ and (b) $\theta$ associated
with the second order standing waves over the CSL structure under finite dc
magnetic field ($\beta_{0}=0.002$). The parameters are taken as $\beta
_{s}=0.02$ and $\beta_{x0}=0.0001$. Analytical result is obtained by Eqs.
(\ref{Split1}) and (\ref{Split2}) with $\epsilon=2/3$ at $\tau=6000$. }%
\label{Fig8}%
\end{figure}

In Fig. \ref{Fig8}, we show comparison between numerical and analytical
results for the spatial profiles of $\varphi$ and $\theta$ associated with the
second order standing waves over the CSL structure under finite dc magnetic field.

We now consider the boundary values $\chi(\bar{z}_{L/2},\tau)$ and $\psi
(\bar{z}_{L/2},\tau)$. To make this in a self-consistent manner, we neglect
the additional Kittel ripple terms in Eqs.(\ref{Split1}) and (\ref{Split2}),
which become vanishingly small in the vicinity of the boundaries,
\begin{align}
\chi(\bar{z},\tau)  &  \sim c_{\chi}\frac{\nu_{q}(\bar{z})}{\nu_{q}(\bar
{z}_{L/2})}\chi(\bar{z}_{L/2},\tau),\label{Approx1}\\
\psi(\bar{z},\tau)  &  \sim c_{\psi}\frac{\nu_{q}(\bar{z})}{\nu_{q}(\bar
{z}_{L/2})}\psi(\bar{z}_{L/2},\tau), \label{Approx2}%
\end{align}
where $c_{\chi}$ and $c_{\psi}$ are some constants.

Substitution of (\ref{Approx1}) and (\ref{Approx2}) into Eqs.(\ref{BC1}) and
(\ref{BC2}) leads to the differential equations for the two unknowns
$\psi\left(  z_{L/2},\tau\right)  $ and $\chi\left(  z_{L/2},\tau\right)  $
\begin{align}
\dot{\psi}\left(  z_{L/2},\tau\right)   &  =-\frac{c_{\chi}}{c_{\psi}}%
A_{q}\chi\left(  z_{L/2},\tau\right)  +\frac{{\beta}_{x}(\tau)}{c_{\psi}}%
\sin\varphi_{0}(z_{L/2}),\label{ODEedge1}\\
\dot{\chi}\left(  z_{L/2},\tau\right)   &  =\frac{c_{\psi}}{c_{\chi}}B_{q}%
\psi\left(  z_{L/2},\tau\right)  , \label{ODEedge2}%
\end{align}
where
\begin{align}
A_{q}  &  =-\frac{\nu_{q}^{^{\prime}}(z_{L/2})}{\nu_{q}(z_{L/2})}+\frac{1}%
{2}\frac{\nu_{q}^{^{\prime\prime}}(z_{L/2})}{\nu_{q}(z_{L/2})}+\frac{D}%
{J}\frac{\nu_{q}^{^{\prime}}(z_{L/2})}{\nu_{q}(z_{L/2})}\varphi_{0}^{^{\prime
}}(z_{L/2})\nonumber\\
&  -\left(  \beta_{\text{s}}+\beta_{0}\right)  \cos\varphi_{0}(z_{L/2}),\\
B_{q}  &  =-\frac{\nu_{q}^{^{\prime}}(z_{L/2})}{\nu_{q}(z_{L/2})}+\frac{1}%
{2}\frac{\nu_{q}^{^{\prime\prime}}(z_{L/2})}{\nu_{q}(z_{L/2})}+\frac{1}%
{2}{\varphi^{^{\prime}}}_{0}^{2}(z_{L/2})\nonumber\\
&  +\frac{D}{J}\varphi_{0}^{^{\prime}}(z_{L/2})-\frac{D}{2J}\varphi
_{0}^{^{\prime\prime}}(z_{L/2})\nonumber\\
&  -\left(  \beta_{\text{s}}+\beta_{0}\right)  \cos\varphi_{0}(z_{L/2}).
\end{align}
Then, the resonance frequency for the end surface spins is
\begin{equation}
\Omega_{\text{surface-Pincus}}=\sqrt{A_{q}B_{q}},
\end{equation}
which is identified with as `surface Pincus' mode. Choosing the initial values
$\psi\left(  z_{L/2},0\right)  =0$ and $\chi\left(  z_{L/2},0\right)  =0$ that
is consistent with the field ${\beta}_{x}(\tau)=\beta_{x0}\sin\left(
\Omega\tau\right)  $, we find
\begin{align}
\chi\left(  z_{L/2},\tau\right)   &  =\frac{\beta_{x0}B_{q}}{c_{\chi}}%
\frac{\sin\varphi_{0}(z_{L/2})}{\Omega_{0}\left(  \Omega_{0}^{2}-\Omega
^{2}\right)  }\left[  \Omega_{0}\sin\left(  \Omega\tau\right)  -\Omega
\sin\left(  \Omega_{0}\tau\right)  \right]  ,\label{PsiTime}\\
\psi\left(  z_{L/2},\tau\right)   &  =\frac{\beta_{x0}}{c_{\psi}}\frac
{\Omega\sin\varphi_{0}(z_{L/2})}{\Omega^{2}-\Omega_{0}^{2}}\left[  \cos\left(
\Omega_{0}\tau\right)  -\cos\left(  \Omega\tau\right)  \right]  ,
\label{ChiTime}%
\end{align}
where $\Omega_{0}=\Omega_{\text{surface-Pincus}}=\Omega
_{\text{interior-Pincus}}$. This matching condition leads to the determination
of the wavenumber $q$.

\begin{figure}[t]
\vspace{10mm}
\begin{center}
\includegraphics[width=70mm,bb=0 0 337 460]{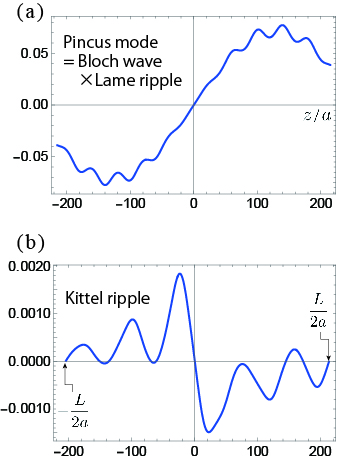}
\end{center}
\caption{Decomposition of Eq. (\ref{Split1}) into the (a) Pincus mode (which
is further decomposed into the Bloch wave and Lame ripple, as shown in Fig.
\ref{Fig6b}) and (b) Kittel ripples.}%
\label{PincusKittel}%
\end{figure}

Analytical results obtained here enable us to understand how the SSWs\ are
constructed from the Bloch waves, Lame ripples and the Kittel ripples. In Fig.
\ref{PincusKittel}, we separately show the spatial profile of $\chi(\bar
{z}_{L/2},\tau)$ and $\epsilon\tilde{\chi}(\bar{z},\tau)$ in Eq.
(\ref{Split1}). The Pincus modes consist of the slowly-varying Bloch wave and
short-wavelength Lam\`{e} ripple. On the other hand, Kittel ripple is
completely pinned down at both ends and has a tiny amplitude.

\begin{figure}[t]
\vspace{10mm}
\begin{center}
\includegraphics[width=70mm,bb=0 0 369 572]{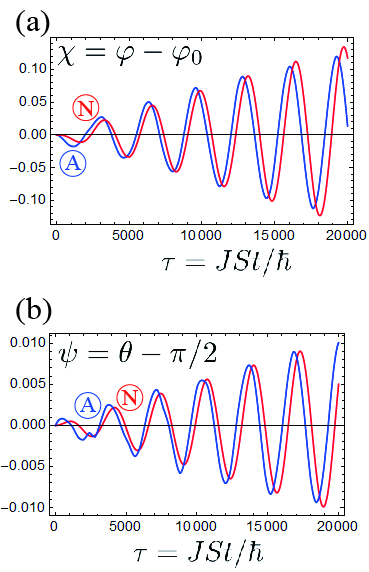}
\end{center}
\caption{Time evolution of $\chi(L/2,\tau)$ and $\psi(L/2,\tau)$: numerical
data (blue) and the analytical relation (\ref{ChiTime}) and (\ref{PsiTime}),
respectively, (red) with $c_{\chi}=60$ and $c_{\psi}=-32$.}%
\label{time-evolution}%
\end{figure}

To make supplementary comparison between numerical and analytical results, in
Fig. \ref{time-evolution} we show the time evolution of $\chi(L/2,\tau)$ and
$\psi(L/2,\tau)$. The analytical results are given by Eqs.(\ref{ChiTime}) and
(\ref{PsiTime}).

\begin{figure}[t]
\vspace{10mm}
\begin{center}
\includegraphics[width=75mm,bb=0 0 390 693]{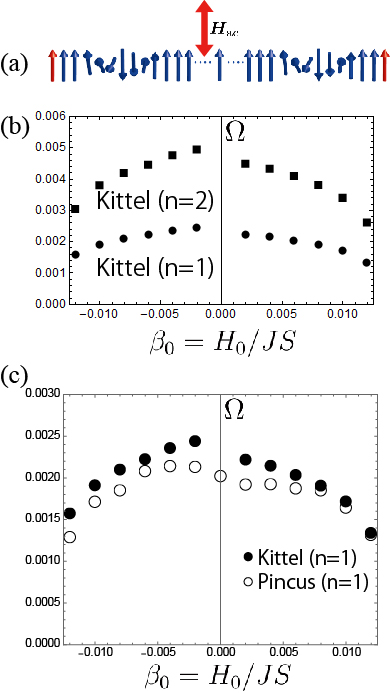}
\end{center}
\caption{(a) Schematic view of the case II configuration. (b) Resonance
frequencies for the Kittel ripples of the lowest orders $n=1$ and $n=2$ shown
by $\CIRCLE$ and $\blacksquare$, respectively. (c) $H_{0}$-dependeces of the
lowest antisymmetric ( $\Circle$ ) and the $n=1$ ripple ( $\CIRCLE$) modes.}%
\label{CaseII-Resonance}%
\end{figure}

Finally, computation of $\tilde{\chi}$ and $\tilde{\psi}$ with the aid of the
found solutions gives resonances of the Kittel ripples at $\Omega_{n}$. In
Fig. \ref{CaseII-Resonance}, we summarize the results which are most essential
results in this paper to make comparison between theoretical and experimental
findings on the magnetic resonance in the case II.

In Fig. \ref{CaseII-Resonance}(c), it is seen that there appear double
resonances. The lower branch originates from the Pincus mode and the upper
branch originates from the Kittel ripple. Actually, the first and second
resonance frequencies are reported as $\Omega_{1}\sim17$GHz and $\Omega
_{2}\sim20$GHz with the bell-shaped field dependence, i.e., $(\Omega
_{2}-\Omega_{1})/\Omega_{1}$ is the order of $0.1$. Based on this fact, we may
conclude that the experimentally observed double resonances consistently
correspond to the first Pincus mode and the first Kittel ripple mode,
respectively. The difference in the resonance frequencies between the first
and second Pincus modes [shown in Fig. \ref{CaseII-Resonance}(b)] is too large
as compared with the experimental finding.

We also note that the intensities of the Pincus mode and Kittel ripples are
almost similar, because they are both the fundamental modes. Based on these
considerations, we may conclude that the resonance profile in the case II is
attributed to the coexistence of the Pincus and Kittel excitations. In the
final part ot Appendix C, we explained the reason why the resonance frequency
for the Pincus mode is smaller than that of the Kittel ripple.

We also recognize an asymmetric profile of the distribution of the resonance
frequencies with respect to the direction of the dc field ($\beta_{0}>0$ or
$\beta_{0}<0$). This asymmetry is actually experimentally
observed.\cite{Goncalves2017} The origin of this asymmetry is easily
understood in terms of the uniform direction of the pinning field at both
ends. Because the pinning fields $\boldsymbol{H}_{s}$ are uniform, the total
field, $\boldsymbol{H}_{0}+\boldsymbol{H}_{s}$, exhibits an asymmetric profile
depending on the orientation of $\boldsymbol{H}_{0}$, i.e., the
$\boldsymbol{H}_{0}$\ is either parallel to or antiparallel to the
$\boldsymbol{H}_{s}$.

\subsubsection{SSW over the simple helix}

As similar to the case I, the case of simple helix can be more easily analyzed
than the case of the CSL. In the case II, the SSW\ is antisymmetric with
respect to reflection across the center. This situation reminds us the
standing waves in thin ferromagnetic films with the asymmetric surface pinning
\cite{Puszkarski1973}. We accordingly modify the scheme given by Eqs.
(\ref{Split1}) and (\ref{Split2}) in the form
\begin{align}
\chi(z,\tau)  &  =\frac{\sin(kz)}{\sin(kL/2)}\chi(L/2,\tau)+\varepsilon
\tilde{\chi}(z,\tau),\label{Split1Sp}\\
\psi(z,\tau)  &  =\frac{\sin(kz)}{\sin(kL/2)}\psi(L/2,\tau)+\varepsilon
\tilde{\psi}(z,\tau). \label{Split2Sp}%
\end{align}
Apparently, the boundary values become antisymmetric, $\chi(-L/2,\tau
)=-\chi(L/2,\tau)$ and $\psi(-L/2,\tau)=-\psi(L/2,\tau)$, provided the
short-range parts vanish at the ends.

\begin{figure}[t]
\vspace{10mm}
\begin{center}
\includegraphics[width=70mm,bb=0 0 368 558]{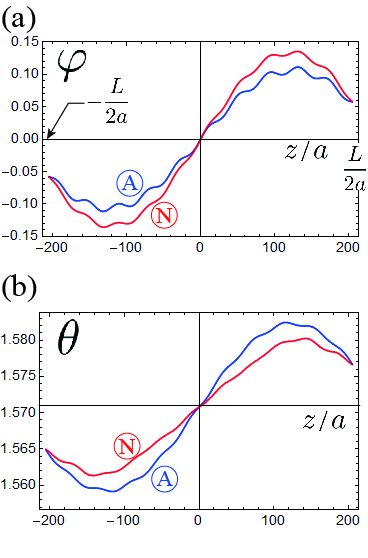}
\end{center}
\caption{ Comparison between numerical (red line) and analytical (blue line)
results for the spatial profiles of (a) $\varphi$ and (b) $\theta$ associated
with the second order standing waves over the helical structure under zero dc
magnetic field ($\beta_{0}=0$). The parameters are taken as $\beta_{\text{s}%
}=0.02$ and $\beta_{x0}=0.0001$. Analytical result is obtained by Eqs.
(\ref{Split1Sp}) and (\ref{Split2Sp}) with $\epsilon=-2/3$ at $\tau=9500$. }%
\label{Fig11}%
\end{figure}

We make use (\ref{Split1Sp}) and (\ref{Split2Sp}) for the system
\begin{align}
\partial_{\tau}\psi &  =-\partial_{z}^{2}\chi+{\beta}_{x}(\tau)\sin(q_{s}z),\\
\partial_{\tau}\chi &  =\left[  \partial_{z}^{2}-2({D}/{J})q_{s}+q_{s}%
^{2}\right]  \psi,
\end{align}
and seek for the solution in the form of the Fourier series
\begin{align}
\tilde{\psi}(z,\tau)  &  =\sum_{n=1}^{\infty}\tilde{\psi}_{n}(\tau)\sin\left(
{2\pi nz}/{L}\right)  ,\label{sinFTfluct2}\\
\tilde{\chi}(z,\tau)  &  =\sum_{n=1}^{\infty}\tilde{\chi}_{n}(\tau)\sin\left(
{2\pi nz}/{L}\right)  . \label{sinFTfluct1}%
\end{align}
Here,
\begin{align}
\tilde{\chi}_{n}(\tau)  &  =f_{n}\frac{\beta_{x0}}{\epsilon}\sqrt{\frac
{q_{n}^{2}-q_{s}^{2}+2q_{s}(D/J)}{q_{n}^{2}}}\nonumber\\
&  \times\frac{\left[  \Omega_{n}\sin\left(  \Omega\tau\right)  -\Omega
\sin\left(  \Omega_{n}\tau\right)  \right]  }{\Omega^{2}-\Omega_{n}^{2}},
\label{SinChi}%
\end{align}%
\begin{equation}
\tilde{\psi}_{n}(\tau)=f_{n}\frac{\beta_{x0}}{\epsilon}\Omega\,\frac{\left[
\cos\left(  \Omega_{n}\tau\right)  -\cos\left(  \Omega\tau\right)  \right]
}{\Omega^{2}-\Omega_{n}^{2}} \label{SinPsi}%
\end{equation}
with the coefficients being given by
\begin{equation}
f_{n}=\frac{2}{L}\int_{-L/2}^{L/2}\sin\left(  q_{s}z\right)  \sin\left(  {2\pi
nz}/{L}\right)  dz.
\end{equation}
The wavevectors are given by $q_{n}=2\pi n/L$ such that $\sin\left(
q_{n}L/2\right)  =0$. The resonance frequency for the Kittel ripple is then
given by
\begin{equation}
\Omega_{n}^{2}=q_{n}^{2}\left[  q_{n}^{2}-q_{s}^{2}+2q_{s}(D/J)\right]  .
\label{DFreq}%
\end{equation}
On the other hand, the resonance frequency for the Pincus mode is obtained
through
\begin{align}
&  \left[  k\cot(kL/2)\left(  1-q_{s}D/J\right)  +k^{2}/2+\beta_{s}%
\cos(kL/2)\right] \nonumber\\
&  \times\left[  k\cot(kL/2)+k^{2}/2-q_{s}^{2}/2\right. \nonumber\\
&  \left.  -(D/J)q_{s}+\beta_{s}\cos(kL/2)\right] \nonumber\\
&  =k^{2}\left[  k^{2}-q_{s}^{2}+2q_{s}(D/J)\right]  .
\end{align}
Numerical estimates with the same parameters as in the previous subsection
give the value $k^{(2)}=0.0126038$. In Fig. \ref{Fig11}, we show comparison
between the numerical and analytical results.

To specify boundary dynamics, we substitute
\begin{equation}
\chi(\bar{z},\tau)\sim c_{\chi}\frac{\sin(kz)}{\sin(kL/2)}\chi(L/2,\tau),
\label{Approx1sin}%
\end{equation}%
\begin{equation}
\psi(z,\tau)\sim c_{\psi}\frac{\sin(kz)}{\sin(kL/2)}\psi(L/2,\tau),
\label{Approx2sin}%
\end{equation}
into Eqs. (\ref{BC1}) and (\ref{BC2}) that yields
\begin{align}
\chi\left(  z_{L/2},\tau\right)   &  =\frac{\beta_{x0}}{c_{\chi}}\frac
{B_{k}\sin\varphi_{0}(z_{L/2})}{\Omega_{0}\left(  \Omega_{0}^{2}-\Omega
^{2}\right)  }\left[  \Omega_{0}\sin\Omega\tau-\Omega\sin\Omega_{0}%
\tau\right]  ,\label{ChiTimeSin}\\
\psi\left(  z_{L/2},\tau\right)   &  =\frac{\beta_{x0}}{c_{\psi}}\frac
{\Omega\sin\varphi_{0}(z_{L/2})}{\left(  \Omega^{2}-\Omega_{0}^{2}\right)
}\left[  \cos\Omega_{0}\tau-\cos\Omega\tau\right]  , \label{PsiTimeSin}%
\end{align}
where
\begin{equation}
B_{k}=-k\cot(kL/2)-k^{2}/2+q_{s}^{2}/2+(D/J)q_{s}-\beta_{\text{s}}\cos(kL/2).
\end{equation}

\begin{figure}[t]
\vspace{10mm}
\begin{center}
\includegraphics[width=70mm,bb=0 0 375 569]{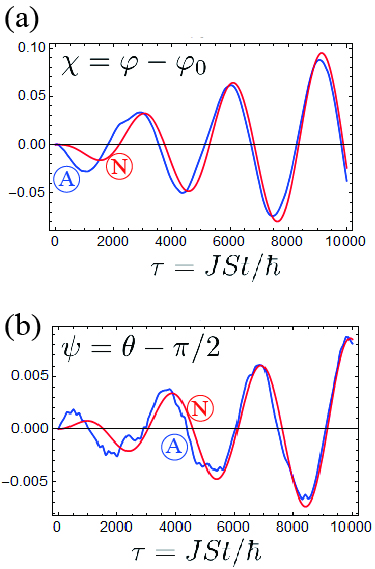}
\end{center}
\caption{Time evolution of $\chi(L/2,\tau)$ and $\psi(L/2,\tau)$: numerical
data (blue) compared to the analytical relations (\ref{ChiTime}) and
(\ref{PsiTime}), respectively, (red) with $c_{\chi}=-12.5$ and $c_{\psi
}=-19.0$. }%
\label{Fig12}%
\end{figure}

These results show that the end spins oscillate with the frequency of
unperturbed standing wave
\begin{equation}
\Omega_{0}^{2}={k}^{2}\left[  k^{2}-q_{s}^{2}+2q_{s}(D/J)\right]  ,
\end{equation}
where the numerical estimation gives $\Omega_{0}\approx0.0020185$. For
comparison, the lowest short-range ripple frequency is $\Omega_{1}=0.0024617$.
Figs. \ref{Fig12}(a), (b) illustrate comparison of (\ref{ChiTimeSin}) and
(\ref{PsiTimeSin}) with numerical data.

\section{Discussions and conclusions}

In this paper, motivated by experimental findings on the magnetic
field-dependence of the resonance profile in CrNb$_{3}$S$_{6}$%
,\cite{Goncalves2017} we developed a theory of the SSW in a mono-axial chiral
helimagnet. We assumed that micro-sized samples used in
Ref.\cite{Goncalves2017} are described as thin films with both the surface end
spins being softly pinned and constructed a theory along with the line of
Kittel-Pincus theories of ferromagnetic resonance. From technical viewpoints,
we presented a scheme of the Davis-Puszkarski equation generalized to the case
of chiral helimagnetic structure under the static magnetic field applied
perpendicular to the chiral axis.

Consequently, we found there are two classes of the SSW over the spatially
modulated chiral spin soliton lattice state. One is the soft Pincus mode,
while another is the hard Kittel ripple. The former is anologus to the SSW in
a ferromagnetic thin film discussed by Pincus.\cite{Pincus1960} The latter
appears only in the case of spatially modulated spin structure, because the
spatially oscillating field acts on the interior spins and causes forced
oscillation in space-domain. The Kittel ripples are excited only when the ac
magnetic field is applied perpendicular to the chiral axis, because the
perpendicular field couples with the spatially modulated component of the
spins through the term $\beta_{x}(t)$cos$\varphi_{0}\left(  z\right)  $. The
existence of two modes, Pincus modes and Kitttel ripples are consistent with
the experimentally observed double resonance profile.

Standing spin waves in thin ferromagnetic films have been intensively studied
from both theoretical and experimental standpoints. These excitations have
already found a broad application in problems of spintronics. However, to best
of our knowledge, problem on the SSW for noncolinear magnetic structure has
not been addressed so far. The present paper may open a new direction in the field.

Finally, we make some general comments on the issues which have not been
treated in this paper. Generally speaking, consideration on the chiral
helimagnets requires careful treatment of dipole-exchange and magneto-static
modes, which play an essential role in films of thickness typically in the
range of micrometers.\cite{Damon1961,Hurben1995,Arias2016} Another important
aspect involved in practical application of the standing spin waves pertains
to accounting for damping effects not dealt with in this work. The theoretical
studies for thin films of itinerant ferromagnet have shown that the Landau
damping mechanism of the standing wave modes is sufficiently severe, since
they have a finite and rather large vector normal to the film
surfaces.\cite{Costa2004} An analysis of the perpendicular standing waves in
sputtered permalloy films by means of a waveguide based FMR measurements
identifies three contributions to their damping: the intrinsic damping, the
eddy-current damping,\cite{Lock1966} and the radiative damping that stem from
the inductive coupling between sample and the waveguide.\cite{Schoen2015} The
radiative damping presents in all ferromagnets including insulators.

The eddy-current damping may be materialized for standing waves in CrNb$_{3}%
$S$_{6}$ films, since there are itinerant carriers from conducting NbS$_{2}$
layers. However, even studies in itinerant ferromagnets, quite thick permalloy
films, demonstrate that the eddy-current damping may be neglected in higher
order standing waves. In addition, in thin ferromagnetic films the
eddy-current damping was found to be negligible in comparison with the
wave-number-dependent damping mechanism due to intra-layer spin-current
transport.\cite{Li2016} Further theoretical and experimental exploration
toward this direction is of interest for future investigations.

\begin{acknowledgments}
The authors thank Y. Kato and J. Ohe for fruitful discussions. This work was
supported by a Grant-in-Aid for Scientific Research (B) (No. 17H02923) from
the MEXT of the Japanese Government, JSPS Bilateral Joint Research Projects
(JSPS-FBR), and the JSPS Core-to-Core Program, A. Advanced Research Networks.
Vl.E.S. and I.P. acknowledge financial support by the Ministry of Education
and Science of the Russian Federation, Grant No. MK-1731.2018.2 and by the
Russian Foundation for Basic Research (RFBR), Grant No. 18-32-00769 (mol\_a).
A.S.O. and I.G.B. acknowledge funding by the RFBR, Grant No. 17-52-50013, by
the Foundation for the Advancement of Theoretical Physics and Mathematics
BASIS Grant No. 17-11-107, and by Act 211 Government of the Russian
Federation, contract No. 02.A03.21.0006. A.S.O. thanks the Ministry of
Education and Science of the Russian Federation, Project No. 3.2916.2017/4.6.
\end{acknowledgments}

\appendix

\section{Lam\'{e} form of equations of motion}

We here derive Eqs. (\ref{Lame1}) and (\ref{Lame2}). First, in Eqs.
(\ref{EOM1}) and (\ref{EOM2}), we make linear expansion with respect to $\psi$
and $\chi$ to obtain%
\begin{align*}
\sin\theta\partial_{z}^{2}\varphi &  \rightarrow\partial_{z}^{2}\varphi
_{0}+\partial_{z}^{2}\chi,\\
\beta_{0}\sin\varphi &  \rightarrow\beta_{0}\sin\varphi_{0}+\left(  \beta
_{0}\cos\varphi_{0}\right)  \chi,\\
\beta_{x}\sin\varphi &  \rightarrow\beta_{x}\sin\varphi_{0},\\
\sin\theta\cos\theta\left(  \partial_{z}\varphi\right)  ^{2}  &
\rightarrow-\psi\left(  \partial_{z}\varphi_{0}\right)  ^{2},\\
\sin\theta\cos\theta\partial_{z}\varphi &  \rightarrow-\psi\partial_{z}%
\varphi_{0},
\end{align*}
where we note $\varphi_{0}$ satisfies the sine-Gordon equation, $\partial
_{z}^{2}\varphi_{0}=\beta_{0}\sin\varphi_{0}$. The higer order terms such as
$\psi\partial_{z}\psi$, $\beta_{x}\chi$ and $\beta_{x}\psi$ are discarded
($\beta_{x}$ is treated as a small perturbation). Plugging them into Eqs.
(\ref{EOM1}) and (\ref{EOM2}) and introducing dimensionless space and time
variables, $\bar{z}=z\sqrt{\beta_{0}}/\kappa$ and $\tau=JSt/\hbar$, we have%
\begin{align}
\frac{\kappa^{2}}{\beta_{0}}\partial_{\tau}\psi &  =\left(  -\frac
{\partial^{2}}{\partial\bar{z}^{2}}+\kappa^{2}\cos\varphi_{0}\right)
\chi+\frac{\kappa^{2}}{\beta_{0}}\beta_{x}(\tau)\sin\varphi_{0},\label{EOM01}%
\\
\frac{\kappa^{2}}{\beta_{0}}\partial_{\tau}\chi &  =-\left[  -\frac
{\partial^{2}}{\partial\bar{z}^{2}}+\kappa^{2}\cos\varphi_{0}-\left(
{\partial_{\bar{z}}\varphi}_{0}\right)  ^{2}\right. \nonumber\\
&  \left.  +2\frac{D}{J}\frac{\kappa}{\sqrt{\beta_{0}}}\left(  {\partial
_{\bar{z}}}\varphi_{0}\right)  \right]  \psi-\frac{\kappa^{2}}{\beta_{0}}%
\beta_{z}(\tau). \label{EOM02}%
\end{align}
Using Eq. (\ref{Ground}), it may be shown%
\begin{align}
\cos\varphi_{0}  &  =-1+2\text{sn}^{2}\bar{z},\\
{\partial_{\bar{z}}}\varphi_{0}  &  =2\text{dn}\bar{z}.
\end{align}
and Eq. (\ref{EOM01}) is reduced to Eq. (\ref{Lame1}). For small $\kappa$, we
can approximate%
\begin{align}
\frac{\kappa}{\sqrt{\beta_{0}}}  &  =\frac{4JE(\kappa)}{\pi D}\sim\frac{2J}%
{D}\left(  1-\frac{\kappa^{2}}{4}\right)  ,\\
\text{dn}\bar{z}  &  =\sqrt{1-\kappa^{2}\text{sn}^{2}\bar{z}}\sim
1-\frac{\kappa^{2}}{2}\text{sn}^{2}\bar{z}.
\end{align}
Then, Eq. (\ref{EOM02}) is reduced to%
\begin{align}
\frac{\kappa^{2}}{\beta_{0}}\partial_{\tau}\chi &  =-\left(  -\frac
{\partial^{2}}{\partial\bar{z}^{2}}+2\kappa^{2}\text{sn}^{2}\bar{z}%
-3\kappa^{2}+4\right)  \psi\nonumber\\
&  -\frac{\kappa^{2}}{\beta_{0}}\beta_{z}(\tau),
\end{align}
which gives Eq. (\ref{Lame2}).

\section{Dynamics of the surface end spins}

We examine in detail the motion of the end spins with the coordinates $\pm
L/2$ in a line of $N$ spins along the $z$ axis. It is assumed that the end
spins experience an effective 'surface' anisotropy field $H_{\text{s}}$, the
same on both ends, which is perpendicular to the line and to the static
magnetic field $H_{0}$.

The boundary spins differ from interior ones in that they have only one
nearest neighbour instead of the usual two, thereby giving the Hamiltonian for
the right end spin
\begin{align}
\mathcal{H}_{R}  &  =-J\boldsymbol{S}_{L/2-1}\cdot\boldsymbol{S}%
_{L/2}-D\left[  \boldsymbol{S}_{L/2-1}\times\boldsymbol{S}_{L/2}\right]
_{z}\nonumber\\
&  -\left(  H_{0}+H_{\text{s}}\right)  S_{L/2}^{x}. \label{HamBC}%
\end{align}

The equations of motion are then shaped into
\begin{align}
\partial_{\tau}\theta_{L/2}  &  =\sin\theta_{L/2-1}\sin\left(  \varphi
_{L/2}-\varphi_{L/2-1}\right) \nonumber\\
&  -\left(  D/J\right)  \sin\theta_{L/2-1}\cos\left(  \varphi_{L/2}%
-\varphi_{L/2-1}\right) \nonumber\\
&  +\left(  \beta_{0}+\beta_{\text{s}}\right)  \sin\varphi_{L/2},
\label{BCdisc1}%
\end{align}%
\begin{gather}
\sin\theta_{L/2}\partial_{\tau}\varphi_{L/2}=\cos\theta_{L/2}\sin
\theta_{L/2-1}\cos\left(  \varphi_{L/2-1}-\varphi_{L/2}\right) \nonumber\\
-\sin\theta_{L/2}\cos\theta_{L/2-1}\nonumber\\
+\left(  D/J\right)  \cos\theta_{L/2}\sin\theta_{L/2-1}\sin\left(
\varphi_{L/2}-\varphi_{L/2-1}\right) \nonumber\\
+\left(  \beta_{0}+\beta_{\text{s}}\right)  \cos\theta_{L/2}\cos\varphi_{L/2}.
\label{BCdisc2}%
\end{gather}
The system (\ref{BC1}) and (\ref{BC2}) originates from (\ref{BCdisc1}) and
(\ref{BCdisc2}) in a continuum limit. Eq.(\ref{stat_right}) for the wave
vector of the ground state is obtained if to take $\varphi_{i}=q_{s}z_{i}$ and
$\theta_{i}=\pi/2$. A treatment for the left edge may be done in a similar way.

\section{General scheme of the Davis-Puszkarski equation for one-dimensional
noncolinear magnetic structure}

For convenience of applications to the standing spin wave problem in a
non-colinear magnetic chain, we here give a concise summary of the
Davis-Puszkarski equation in a self-contained manner. The notation in this
appendix will be independent of that of the main body of the paper.

\subsection{Equations of motion for the interior and the ends}

We represent the arrangement of the spins on a chain with a linear length
$L$\ using the polar coordinates, $\boldsymbol{S}\left(  z\right)  =S\left(
\sin\theta\cos\varphi,\sin\theta\sin\varphi,\cos\theta\right)  $. The interior
system is described by a Lagrangian in a general form,%
\begin{equation}
L\left[  \theta,\varphi\right]  =\hbar S\int_{-\frac{L}{2}}^{\frac{L}{2}%
}\left(  \cos\theta-1\right)  \partial_{t}\varphi dz-\int_{-\frac{L}{2}%
}^{\frac{L}{2}}\mathcal{H}\left[  \theta,\varphi\right]  dz,
\end{equation}
where the first term represents the kinetic Berry phase term. The effective
Hamiltonian $\mathcal{H}\left[  \theta,\varphi\right]  $ describes a
continuous model which contains spatial derivatives of $\theta$ and $\varphi
$\ fields coming from the exchange interactions and non-linear Zeeman terms
such as $\sin\theta\cos\varphi$ or cos$\theta$. Then, the coupled
Euler-Lagrangian equations of motion are written down to be%

\begin{align}
\hbar S\sin\theta\frac{\partial\theta}{\partial t}  &  =\frac{\delta
\mathcal{H}}{\delta\varphi},\label{EOMappendix01}\\
\hbar S\sin\theta\frac{\partial\varphi}{\partial t}  &  =-\frac{\delta
\mathcal{H}}{\delta\theta}. \label{EOMappendix02}%
\end{align}
Next we assume that the ground state configuration $\theta_{0}\left(
z\right)  $\ and $\varphi_{0}\left(  z\right)  $\ are given through the
stationarity condition $\delta\int\mathcal{H}\left[  \theta,\varphi\right]
dz=0$ and consider the small (Gaussian) fluctuations $\delta\theta\left(
z,t\right)  $\ and $\delta\varphi\left(  z,t\right)  $,%
\begin{align}
\theta\left(  z,t\right)   &  =\theta_{0}\left(  z\right)  +\delta
\theta\left(  z,t\right)  ,\\
\varphi\left(  z,t\right)   &  =\varphi_{0}\left(  z\right)  +\delta
\varphi\left(  z,t\right)  .
\end{align}
Expanding the EOMs (\ref{EOMappendix01}) and (\ref{EOMappendix02}), we may
obtain the EOMs for the fluctuations in a general form,%

\begin{align}
\frac{\partial\delta\theta\left(  z,t\right)  }{\partial t}  &  =\hat
{\mathcal{L}}_{\varphi}\delta\varphi\left(  z,t\right)  +\epsilon f\left(
z,t\right)  ,\label{EOMfluc001}\\
\frac{\partial\delta\varphi\left(  z,t\right)  }{\partial t}  &
=-\hat{\mathcal{L}}_{\theta}\delta\theta\left(  z,t\right)  ,
\label{EOMfluc002}%
\end{align}
where $\hat{\mathcal{L}}_{\varphi}$\ and $\hat{\mathcal{L}}_{\theta}$\ are
linear differential operators. Without loss of generality, we introduce a
small external force term $\epsilon f\left(  z,t\right)  $. From now on, we
consider two cases: $\epsilon=0$\ and finite $\epsilon\neq0$.

Next we address to EOMs for the end spins. We start with the lattice
Hamiltonian and write down the EOMs for the end spins, $\boldsymbol{S}_{1}$
and $\boldsymbol{S}_{N}$, which couple with the nearest neighbor interior
spins $\boldsymbol{S}_{2}$ and $\boldsymbol{S}_{N-1}$, respectively. For
example, the right-side end ($z=L/2$), the continuum limit is taken as%
\begin{equation}
\varphi_{_{N-1}}\sim\varphi(L/2)-a\frac{\partial\varphi\left(  z\right)
}{\partial z}+\frac{a^{2}}{2}\frac{\partial^{2}\varphi\left(  z\right)
}{\partial z^{2}},
\end{equation}
and then evaluating the derivative at $z=L/2$. Thus, we obtain the EOMs at the
ends,
\begin{align}
\frac{\partial\delta\theta_{\text{s}}\left(  z,t\right)  }{\partial t}  &
=\hat{\mathcal{M}}_{\varphi}\delta\varphi_{\text{s}}\left(  z,t\right)
+\epsilon f\left(  z,t\right)  ,\label{EOMflucS001}\\
\frac{\partial\delta\varphi_{\text{s}}\left(  z,t\right)  }{\partial t}  &
=-\hat{\mathcal{M}}_{\theta}\delta\theta_{\text{s}}\left(  z,t\right)  .
\label{EOMflucS002}%
\end{align}
Here, $\hat{\mathcal{M}}_{\varphi}$ and $\hat{\mathcal{M}}_{\theta}$ are the
linear operators including the effects of the surface pinning fields. After
acting by these operators on $\delta\varphi_{\text{s}}\left(  z,t\right)
$\ and $\delta\theta_{\text{s}}\left(  z,t\right)  $, respectively, we fix
then $z=z_{\text{s}}=\pm L/2$, and, consequently, a dependence on the interior
coordinate, $z$, disappears.

\subsection{The case of $\epsilon=0$}

In the case of $\epsilon=0$, Eqs. (\ref{EOMfluc001}) and (\ref{EOMfluc002})
are solved using separation of variables,%
\begin{align}
\delta\theta\left(  z,t\right)   &  =\mu\left(  z\right)  M\left(  t\right)
,\\
\delta\varphi\left(  z,t\right)   &  =\nu\left(  z\right)  N\left(  t\right)
.
\end{align}
Inserting these forms to Eqs. (\ref{EOMfluc001}) and (\ref{EOMfluc002}), we
obtain%
\begin{align}
\frac{1}{N\left(  t\right)  }\frac{\partial M\left(  t\right)  }{\partial t}
&  =\frac{1}{\mu\left(  z\right)  }\hat{\mathcal{L}}_{\varphi}\nu\left(
z\right)  =C_{1},\\
\frac{1}{M\left(  t\right)  }\frac{\partial N\left(  t\right)  }{\partial t}
&  =-\frac{1}{\nu\left(  z\right)  }\hat{\mathcal{L}}_{\theta}\mu\left(
z\right)  =-C_{2},
\end{align}
with $C_{1}$\ and $C_{2}$\ being constants. Then the temporal constituents
immediately gives the eigenfrequency for the interior system,
\begin{equation}
\Omega_{\text{interior}}=\sqrt{C_{1}C_{2}}. \label{AppendixIneriorOmega01}%
\end{equation}

Next let us consider the spatial parts,%
\begin{align}
\hat{\mathcal{L}}_{\varphi}\nu\left(  z\right)   &  =C_{1}\mu\left(  z\right)
,\\
\hat{\mathcal{L}}_{\theta}\mu\left(  z\right)   &  =C_{2}\nu\left(  z\right)
.
\end{align}
Here we assume that the differential operators $\hat{\mathcal{L}}_{\theta}$
and $\hat{\mathcal{L}}_{\varphi}$\ have simultaneous eigenfunctions, $\Psi
_{q}\left(  z\right)  $, labeled by an index $q$. That is to say,%
\begin{align}
\hat{\mathcal{L}}_{\theta}\Psi_{q}\left(  z\right)   &  =\lambda_{\theta
}\left(  q\right)  \Psi_{q}\left(  z\right)  ,\\
\hat{\mathcal{L}}_{\varphi}\Psi_{q}\left(  z\right)   &  =\lambda_{\varphi
}\left(  q\right)  \Psi_{q}\left(  z\right)  ..
\end{align}
Expanding $\mu\left(  z\right)  $ and $\nu\left(  z\right)  \ $in terms of
$\Psi_{q}\left(  z\right)  $ as the orthogonal basis,%
\begin{align}
\mu\left(  z\right)   &  =%
%TCIMACRO{\tsum _{q}}%
%BeginExpansion
{\textstyle\sum_{q}}
%EndExpansion
\mu_{q}\Psi_{q}\left(  z\right)  ,\\
\nu\left(  z\right)   &  =%
%TCIMACRO{\tsum _{q}}%
%BeginExpansion
{\textstyle\sum_{q}}
%EndExpansion
\nu_{q}\Psi_{q}\left(  z\right)  ,
\end{align}
we have%
\begin{align}%
%TCIMACRO{\tsum _{q}}%
%BeginExpansion
{\textstyle\sum_{q}}
%EndExpansion
\left[  \nu_{q}\lambda_{\varphi}\left(  q\right)  -C_{1}\mu_{q}\right]
\Psi_{q}\left(  z\right)   &  =0,\\%
%TCIMACRO{\tsum _{q}}%
%BeginExpansion
{\textstyle\sum_{q}}
%EndExpansion
\left[  \mu_{q}\lambda_{\theta}\left(  q\right)  -C_{2}\nu_{q}\right]
\Psi_{q}\left(  z\right)   &  =0,
\end{align}
which lead to%
\begin{equation}
C_{1}=\frac{\nu_{q}}{\mu_{q}}\lambda_{\varphi}\left(  q\right)  \text{, }%
C_{2}=\frac{\mu_{q}}{\nu_{q}}\lambda_{\theta}\left(  q\right)  \text{.}%
\end{equation}
Therefore, we can replace Eq. (\ref{AppendixIneriorOmega01}) with%
\begin{equation}
\Omega_{\text{interior}}\left(  q\right)  =\sqrt{\lambda_{\theta}\left(
q\right)  \lambda_{\varphi}\left(  q\right)  }.
\end{equation}

Now, let us turn attention to the fluctuations at the ends, described in a
form%
\begin{align}
\delta\theta_{\text{s}}\left(  t\right)   &  =\delta\theta\left(  z_{\text{s}%
},t\right)  =c_{1}\Psi_{q}\left(  z_{\text{s}}\right)  M_{\text{s}}\left(
t\right)  ,\label{surfacetheta01}\\
\delta\varphi_{\text{s}}\left(  t\right)   &  =\delta\varphi\left(
z_{\text{s}},t\right)  =c_{2}\Psi_{q}\left(  z_{\text{s}}\right)  N_{\text{s}%
}\left(  t\right)  , \label{surfacephi01}%
\end{align}
with $c_{1}$ and $c_{2}$ being constants, and $z_{\text{s}}=\pm L/2$. Note
that the surface states `participate in' or `swallowed up by' the interior
mode specified by $q$. Inserting them into Eqs. (\ref{EOMflucS001}) and
(\ref{EOMflucS002}), we may have%
\begin{align}
\frac{\partial M_{\text{s}}\left(  t\right)  }{\partial t}  &  =A_{q}%
N_{\text{s}}\left(  t\right)  ,\label{EOMsurfaceApp01}\\
\frac{\partial N_{\text{s}}\left(  t\right)  }{\partial t}  &  =-B_{q}%
M_{\text{s}}\left(  t\right)  , \label{EOMsurfaceApp02}%
\end{align}
where \ $A_{q}$ and $B_{q}$ contain spatial derivatives of $\Psi_{q}\left(
z\right)  $ at $z=z_{\text{s}}$, i.e., no liner differential operator appears.
We dropped an external force term, since information on $A_{q}$ and $B_{q}$ is
enough to obtain the Larmor frequency of the surface end spins,,
\begin{equation}
\Omega_{\text{surface}}(q)=\sqrt{A_{q}B_{q}}.
\end{equation}
Now, the matching condition
\begin{equation}
\Omega_{n}\equiv\Omega_{\text{surface}}(q_{n})=\Omega_{\text{interior}}(q_{n})
\end{equation}
gives the Davis-Puszkarski equation which determines a series of allowed
values of $q_{n}$, ($n=1,2,\cdots$) corresponding to the SSW modes. In the
context of this paper, this corresponds to the Pincus mode.

\subsection{The case of $\epsilon\neq0$}

The existence of space-time dependent term $\epsilon f\left(  z,t\right)  $
prevents us from using separation of variables. We then seek for the interior
solution in a perturbative manner based on an ansatz,%
\begin{align}
\delta\theta\left(  z,t\right)   &  =\frac{\mu\left(  z\right)  }%
{\mu_{\text{s}}}\delta\theta_{\text{s}}\left(  t\right)  +\epsilon\tilde{\psi
}\left(  z,t\right)  ,\\
\delta\varphi\left(  z,t\right)   &  =\frac{\nu\left(  z\right)  }%
{\nu_{\text{s}}}\delta\varphi_{\text{s}}\left(  t\right)  +\epsilon\tilde
{\chi}\left(  z,t\right)  ,
\end{align}
where $\mu_{\text{s}}$ and $\nu_{\text{s}}$ are respectively the values of
$\mu\left(  z\right)  $ and $\nu\left(  z\right)  $ at the surface ends. In
this ansatz, we implicitly assume that the time-dependence at the surface,
$\delta\theta_{\text{s}}\left(  t\right)  $ and $\delta\varphi_{\text{s}%
}\left(  t\right)  $, are known. We here impose the boundary condition,%
\begin{equation}
\tilde{\psi}\left(  z_{\text{s}},t\right)  =\tilde{\chi}\left(  z_{\text{s}%
},t\right)  =0,
\end{equation}
which means the perfect pinning of the $\tilde{\chi}$\ and $\tilde{\psi}%
$\ fields at the ends. Due to these conditions, we consistently reproduce%
\begin{align}
\delta\theta\left(  z_{\text{s}},t\right)   &  =\delta\theta_{\text{s}}\left(
t\right)  ,\\
\delta\varphi\left(  z_{\text{s}},t\right)   &  =\delta\varphi_{\text{s}%
}\left(  t\right)  ,
\end{align}
at the surfaces.

\subsubsection{Pincus mode}

Now, we proceed with analysis in a perturbative manner. Collecting the zeroth
order terms with respect to $\epsilon$, we have%
\begin{align}
\frac{1}{\delta\varphi_{\text{s}}\left(  t\right)  }\frac{\partial\delta
\theta_{\text{s}}\left(  t\right)  }{\partial t}  &  =\frac{\mu_{\text{s}}%
}{\nu_{\text{s}}}\frac{1}{\mu\left(  z\right)  }\hat{\mathcal{L}}_{\varphi}%
\nu\left(  z\right)  =D_{1},\\
\frac{1}{\delta\theta_{\text{s}}\left(  t\right)  }\frac{\partial\delta
\varphi_{\text{s}}\left(  t\right)  }{\partial t}  &  =-\frac{\nu_{\text{s}}%
}{\mu_{\text{s}}}\frac{1}{\nu\left(  z\right)  }\hat{\mathcal{L}}_{\theta}%
\mu\left(  z\right)  =-D_{2},
\end{align}
with $D_{1}$\ and $D_{2}$\ being constants. Then, as in the case of
$\epsilon=0$, the eigenfrequency for the interior system is given by
\begin{equation}
\Omega_{\text{interior}}\left(  q\right)  =\sqrt{D_{1}D_{2}}=\sqrt
{\lambda_{\varphi}\left(  q\right)  \lambda_{\theta}\left(  q\right)  }.
\end{equation}

Let us then turn attention to the surface spins. In the vicinity of the ends,
$\tilde{\psi}\left(  z,t\right)  $ and $\tilde{\chi}\left(  z,t\right)  $ can
be dropped and we have%
\begin{align}
\delta\theta\left(  z,t\right)   &  =c_{1}\Psi_{q}\left(  z\right)
\delta\theta_{\text{s}}\left(  t\right)  .\\
\delta\varphi\left(  z,t\right)   &  =c_{2}\Psi_{q}\left(  z\right)
\delta\varphi_{\text{s}}\left(  t\right)  ,
\end{align}
with $c_{1}$ and $c_{2}$ being constants. It is to be noted that the basis
function $\Psi_{q}\left(  z\right)  $ is now specified just as in the case of
(\ref{surfacetheta01}) and (\ref{surfacephi01}). Inserting them into Eqs.
(\ref{EOMflucS001}) and (\ref{EOMflucS002}), we obtain the zeroth order
equations,%
\begin{align}
\frac{\partial\delta\theta_{\text{s}}\left(  t\right)  }{\partial t}  &
=\frac{c_{2}}{c_{1}}F_{q}\delta\varphi_{\text{s}}\left(  t\right)  ,\\
\frac{\partial\delta\varphi_{\text{s}}\left(  t\right)  }{\partial t}  &
=-\frac{c_{1}}{c_{2}}G_{q}\delta\theta_{\text{s}}\left(  t\right)  ,
\end{align}
where $F_{q}$ and $G_{q}$ contain spatial derivatives of $\Psi_{q}\left(
z\right)  $ at $z=z_{\text{s}}$. We thus obtain the Larmor frequency of the
surface end spins,%
\begin{equation}
\Omega_{\text{surface}}\left(  q\right)  =\sqrt{F_{q}G_{q}}.
\end{equation}
Now, again the matching condition
\begin{equation}
\Omega_{n}\equiv\Omega_{\text{surface}}\left(  q_{n}\right)  =\Omega
_{\text{interior}}\left(  q_{n}\right)  \label{DP-for-Pincus}%
\end{equation}
again gives the Davis-Puszkarski equation. In the context of this paper, this
corresponds to the Pincus mode.

\subsubsection{Kittel ripple}

Next, we consider the first order terms to obtain%
\begin{align}
\frac{\partial\tilde{\psi}\left(  z,t\right)  }{\partial t}  &  =\hat
{\mathcal{L}}_{\varphi}\tilde{\chi}\left(  z,t\right)  +f\left(  z,t\right)
,\label{AppendixEOM01}\\
\frac{\partial\tilde{\chi}\left(  z,t\right)  }{\partial t}  &  =-\hat
{\mathcal{L}}_{\theta}\tilde{\psi}\left(  z,t\right)  , \label{AppendixEOM02}%
\end{align}
where the linear operators, $\hat{\mathcal{L}}_{\varphi}$\ and $\hat
{\mathcal{L}}_{\theta}$, for the interior system appear. Again, we expand
$\tilde{\psi}\left(  z,t\right)  $ and $\tilde{\chi}\left(  z,t\right)  \ $in
terms of $\Psi_{Q}\left(  z\right)  $,%
\begin{align}
\tilde{\psi}\left(  z,t\right)   &  =%
%TCIMACRO{\tsum _{Q}}%
%BeginExpansion
{\textstyle\sum_{Q}}
%EndExpansion
\tilde{\psi}_{Q}\left(  t\right)  \Psi_{Q}\left(  z\right)
,\label{KittelExpM01}\\
\tilde{\chi}\left(  z,t\right)   &  =%
%TCIMACRO{\tsum _{Q}}%
%BeginExpansion
{\textstyle\sum_{Q}}
%EndExpansion
\tilde{\chi}_{Q}\left(  t\right)  \Psi_{Q}\left(  z\right)  .
\label{KittelExpM02}%
\end{align}
In this case, \textit{the allowed }$Q$\textit{ is determined solely by the
perfect pinning condition}%
\begin{equation}
\Psi_{Q_{n}}\left(  z_{\text{s}}\right)  =0. \label{KittelBC}%
\end{equation}
We should note the essential difference between the conditions
(\ref{DP-for-Pincus}) and (\ref{KittelBC}).

We insert Eqs. (\ref{KittelExpM01}) and (\ref{KittelExpM02}) into Eqs.
(\ref{AppendixEOM01}) and (\ref{AppendixEOM02}), take one more time
derivative, multiply the both sides by $\Psi_{Q}\left(  z\right)  $, and
integrate over $z$ to obtain%

\begin{align}
\frac{d^{2}\tilde{\psi}_{Q}\left(  t\right)  }{dt^{2}}  &  =-\lambda_{\theta
}\left(  Q\right)  \lambda_{\varphi}\left(  Q\right)  \tilde{\psi}_{Q}\left(
t\right)  +\frac{df_{Q}\left(  t\right)  }{dt},\label{EOMKittelS01}\\
\frac{d^{2}\tilde{\chi}_{Q}\left(  t\right)  }{dt^{2}}  &  =-\lambda_{\varphi
}\left(  Q\right)  \lambda_{\theta}\left(  Q\right)  \tilde{\chi}_{Q}\left(
t\right)  -\lambda_{\theta}\left(  Q\right)  f_{Q}\left(  t\right)  ,
\label{EOMKittelS02}%
\end{align}
where%
\begin{equation}
f_{Q}\left(  t\right)  =\frac{\int_{-L/2}^{L/2}\Psi_{Q}\left(  z\right)
f\left(  z,t\right)  dz}{\int_{-L/2}^{L/2}\Psi_{Q}^{2}\left(  z\right)  dz}.
\end{equation}
Now the the eigenfrequency,
\begin{equation}
\tilde{\Omega}_{n}=\sqrt{\lambda_{\theta}\left(  Q_{n}\right)  \lambda
_{\varphi}\left(  Q_{n}\right)  }, \label{DP-for-Kittel}%
\end{equation}
specifies the modes associates with $\tilde{\psi}\left(  z,t\right)  $\ and
$\tilde{\chi}\left(  z,t\right)  ,$ i.e., the Kittel ripple in the present
context. Eqs. (\ref{EOMKittelS01}) and (\ref{EOMKittelS02}) are readily solved
to give%
\begin{align}
\tilde{\psi}_{Q_{n}}\left(  t\right)   &  =\frac{1}{\tilde{\Omega}_{n}}%
\int_{0}^{t}\frac{df_{Q}\left(  t^{\prime}\right)  }{dt^{\prime}}\sin
[\tilde{\Omega}_{n}(t-t^{\prime})]dt^{\prime},\\
\tilde{\chi}_{Q_{n}}\left(  t\right)   &  =-\frac{\tilde{\lambda}_{\theta
}\left(  Q\right)  }{\tilde{\Omega}_{n}}\int_{0}^{t}f_{Q}\left(  t^{\prime
}\right)  \sin[\tilde{\Omega}_{n}(t-t^{\prime})]dt^{\prime},
\end{align}
provided $f_{Q}(0)=0$. Finally, we note
\begin{equation}
\Omega_{n}<\tilde{\Omega}_{n},
\end{equation}
because the Kittel modes, $\tilde{\psi}\left(  z,t\right)  $\ and $\tilde
{\chi}\left(  z,t\right)  $, are strictly confined into the system over the
region $-L/2\leq z\leq L/2$. On the other hand, the Pincus mode can be
extended beyond this region. This makes $\Omega_{n}$ smaller than
$\tilde{\Omega}_{n}$ for a common $n$.


\begin{thebibliography}{99}
\bibitem {Termin}We use Kittel's terminology here, where the SSW order
corresponds to a number of half-wavelengths on a line.
\end{thebibliography}
\end{document}